\documentclass[useAMS,usenatbib,letterpaper]{mn2e}
\usepackage{epsfig}

\newcommand{\ud}{{\rmn{d}}}
\newcommand{\vnhat}{\hat{\bmath{n}}}

\title[Fast polarization power spectrum estimation]
{Fast estimation of polarization power spectra using correlation functions}
\author[G.~Chon et al.]
{Gayoung Chon,$^1$\thanks{E-mail: gchon@mrao.cam.ac.uk}
Anthony Challinor,$^1$
Simon Prunet,$^2$
Eric Hivon,$^3$
\newauthor
and Istv\'{a}n Szapudi$^4$
\\
$^1$Astrophysics Group, Cavendish Laboratory, Madingley Road,
Cambridge CB3 0HE, U.K. \\
$^2$Institut d'Astrophysique de Paris, 98bis, Boulevard Arago, 75014 Paris,
France \\
$^3$Infrared Processing and Analysis Center, Caltech, 770 South Wilson Avenue,
Pasadena, CA 91125, U.S.A. \\
$^4$Institute for Astronomy, University of Hawaii, 2680 Woodlawn Dr,
Honolulu, HI 96822, U.S.A.
}

\pagerange{\pageref{firstpage}--\pageref{lastpage}}
\pubyear{2003}

\label{firstpage}

\begin{document}

\maketitle

\begin{abstract}
We present a fast method for estimating the cosmic microwave background
polarization power spectra using unbiased estimates of
heuristically-weighted correlation functions. This extends the
$O(N_\rmn{pix}^{3/2})$ method of~\citet{szapudi01} to polarized data.
If the sky coverage allows the correlation functions to be estimated over the
full range of angular separations, they can be inverted
directly with integral transforms and clean separation of the electric ($E$)
and magnetic ($B$) modes of polarization is obtained exactly in the mean.
We assess the level of $E$-$B$ mixing that arises from apodized integral
transforms when the correlation function can only be estimated for a subset of
angular scales, and show that it is significant for small-area observations.
We introduce new estimators to deal with this case on the spherical sky
that preserve $E$-$B$ separation; their construction requires an additional
integration of the correlation functions but the computational cost is
negligible. We illustrate our methods with application to a large-area
survey with parameters similar to \emph{Planck}, and the small-area
Background Imaging of Cosmic Extragalactic Polarization experiment.
In both cases we show that the errors on the recovered power
spectra are close to theoretical expectations.
\end{abstract}


\section{Introduction}
\label{sec:intro}

With the recent detection of polarization in the cosmic microwave
background (CMB) by the Degree Angular Scale Interferometer
(DASI; \citealt{kovac02}), and the detection of the temperature-polarization
cross correlation by the
\emph{Wilkinson Microwave Anisotropy Probe}
(\emph{WMAP})\footnote{http://map.gsfc.nasa.gov/}~\citep{kogut03},
the immediate goal for upcoming polarization
experiments is to map accurately the polarization power spectra
over a wide range of angular scales.
CMB polarization is generated at last scattering from the local quadrupole
moment of the photon total intensity. The expected r.m.s.\ polarization
is $\sim 6.4\, \mu\rmn{K}$, peaking around an angular scale of
$\sim 10\, \rmn{arcmin}$ (the angle subtended by the width of the last
scattering surface),
making mapping a challenging prospect. Further scattering once the Universe
reionizes tends to destroy polarization on scales that are then sub-Hubble,
but generates additional large-angle polarization~\citep{zaldarriaga97}.
The potential cosmological returns from polarization observations are high.
Current polarization data~\citep{kovac02,kogut03} already allows a
stringent test to be made of the paradigm for structure formation from
initially super-Hubble, passive, adiabatic fluctuations. Furthermore,
the detection by \emph{WMAP} of large-scale power in the
temperature-polarization cross power spectrum has provided new constraints
on the reionization process, and helped lift major degeneracies that
affect the determination of cosmological parameters from the CMB temperature
anisotropies. (In particular, the degeneracies between reionization
optical depth and the scalar spectral index and gravitational wave amplitude.)
Potential returns from future, more precise, observations
include (i) detection of the clean signature of a stochastic background of
gravitational waves~\citep*{seljak97,kamionkowski97}
and hence fine selection between inflation models~\citep{kinney98};
and (ii) evidence for
weak gravitational lensing through the distortion of CMB polarization
on small scales~\citep*{zaldarriaga98,benabed01}.
Estimating the polarization power spectra is an important
intermediate step in achieving these science goals. This paper
addresses that problem, presenting a fast, robust method for estimating
the power spectra from large datasets in the presence of real-world
complications, such as incomplete sky coverage and inhomogeneous noise.

The accurate analysis of CMB data places strong demands on the statistical
methods employed. Even for
total intensity data, which is simpler than polarization,
the extraction of the power spectrum from
upcoming mega-pixel datasets with standard maximum-likelihood
methods (e.g.~\citealt*{bond98}) is beyond the range of any supercomputer.
[The operations count scales as the number of pixels cubed, $N_\rmn{pix}^3$,
while the storage requirements are $O(N_\rmn{pix}^2)$.] In the
search for fast alternatives to brute-force maximum-likelihood power
spectrum estimation, two broad approaches have emerged. In the first,
experiment-specific symmetries are exploited to make the
brute-force analysis tractable, or, if the symmetries are only approximate,
to pre-condition an iterative solution to the likelihood maximisation.
An example of the former is the ingenious ``ring-torus'' method
of~\citet{wandelt01}, while the latter was pioneered by~\citet*{oh99}
during the development of the pipeline for the \emph{WMAP} satellite.
The second class
of methods sacrifice optimality in favour of speed by adopting a more
heuristic weighting of the data (such as inverse weighting with the
noise variance). An estimate of the underlying power spectrum is then
obtained from the raw, rotationally-invariant power spectrum of the
weighted map (the pseudo-$C_l$s; \citealt*{wandelt01b}) either by a direct
linear inversion~\citep*{szapudi01,hivon02} or with likelihood
methods (Wandelt et al.\ 2001; \citealt*{hansen02}).
The linear inversion, which yields
estimators quadratic in the data, can be performed
directly in harmonic space~\citep{hivon02}, or, more simply, by
first transforming to real space (i.e.\ by constructing the correlation
function) and then recovering the power spectrum with a (suitably
apodized) integral transform~\citep{szapudi01}. The estimation of polarization
power spectra is less well explored than for total intensity,
although all of the above methods can, in principle, be extended to handle
polarization. To date, only brute-force maximum
likelihood~\citep[][Munshi et al. in preparation]{kovac02},
minimum-variance quadratic estimators~\citep{tegmark02}, pseudo-$C_l$
methods with statistical~\citep{hansen02b} or direct inversion~\citep{kogut03}
in harmonic space, and real-space correlation function
methods~\citep{sbarra03} have
been demonstrated on polarized data. Of these, only the
pseudo-$C_l$ methods are fast enough
to apply many times (e.g.\ in Monte-Carlo simulations) to mega-pixel maps.
In this paper we extend the fast
correlation-function approach of~\citet{szapudi01} to polarized data.

A new problem that arises when analysing polarized data, that
is absent for total intensity, is the decomposition of the polarized field
into its electric ($E$; sometimes denoted gradient) and magnetic ($B$;
alternatively curl)
components. The scientific importance
of this decomposition is that primordial magnetic polarization is not
generated by density perturbations, and so in standard models is sourced
only by gravitational waves~\citep{seljak97,kamionkowski97}. However,
with incomplete sky coverage, separating the polarization field into
electric and magnetic components is no longer straightforward. Exquisite
monitoring of leakage between $E$ and $B$ in analysis pipelines will be
required if primordial $B$ polarization is to be detected down to the
fundamental confusion limit set by cosmic shear~\citep*{kesden02}.
The question of
performing the $E$-$B$ separation on an incomplete sky has received
considerable attention
recently~\citep*{zaldarriaga01,lewis02,bunn02,chiueh02,bunn02b}.
Methods are now available
for extracting pure measures of the $E$ and $B$ \emph{fields} which can then be
used for subsequent power spectrum estimation. An alternative approach is
to perform a joint (i.e.\ $E$ and $B$) power spectrum analysis of the original
polarization data, removing the need for an additional stage in the analysis
pipeline and the non-optimalities that this may introduce. The
efficacy of maximum-likelihood methods for performing the $E$-$B$ separation
is explored by Munshi et al.\ (in preparation). However, the computational
demands of likelihood methods, and the difficulty in monitoring $E$-$B$ leakage
in a non-linear analysis, motivates the development of fast, unbiased
methods. The correlation-function based approach we develop here, motivated
by~\cite{crittenden02}, has a significant feature in that, with a little
post-processing of the correlation functions, leakage between $E$ and $B$ can
be eliminated in the mean, even for observations covering only a small part of
the sky. The separation is exact in the mean.

The outline of this paper is as follows. In Sec.~\ref{correlation}
we review the polarization functions on the sphere and their relation with
the power spectra. Section~\ref{estimators} presents a fast,
$O(N_\rmn{pix}^{3/2})$ method for computing unbiased estimates of the
correlation functions allowing for heuristic weighting of the data, and
describes power spectrum recovery for large-area surveys where the
correlation function can be estimated for the full range of angular
separations. We illustrate our methods by applying them to a survey
mission with similar parameters to those for
\emph{Planck}\footnote{http://sci.esa.int/home/planck/}.
In Sec.~\ref{windows} we provide a careful analysis of the effect of
incomplete coverage of the correlation functions on the direct extraction
of the power spectra with apodized integral transforms. By constructing
the relevant window functions for small-area observations we show that
leakage from $E$ to $B$ can be a significant problem. We remedy this
deficiency of the method in Sec.~\ref{sec:eb}, where we construct
functions from integrals of the original correlation functions that
contain signal contributions from only $E$ or $B$ in the mean. These
functions can be safely inverted with apodized integral transforms to obtain
properly separated estimates of the $E$ and $B$ power spectra. We apply
this new estimator to a model of the Background Imaging of Cosmic Extragalactic
Polarization experiment
(BICEP)\footnote{http://bicep.caltech.edu} experiment,
and show that it produces error bars close to the theoretical expectations.
Our conclusions are given in Sec.~\ref{conclusion}, and the Appendix
contains some technical results on the analytic normalisation of the
correlation functions estimators for uniform weighting on
azimuthally-symmetric patches.

Throughout this paper we illustrate our results with a flat, $\Lambda$CDM model
with concordance parameters $\Omega_\rmn{b} h^2 = 0.022$, $\Omega_\rmn{c}h^2=
0.12$, $\Omega_\Lambda = 0.7$ giving the Hubble constant $h=0.69$. The
primordial scalar curvature and tensor spectra are scale-invariant and have
ratio $r=0.31$ (making the ratio of tensor to scalar power in temperature
anisotropies $r_{10}=0.16$ at $l=10$)\footnote{%
We define $r$ as the ratio of the amplitudes $A_S$ and $A_T$ of the
primordial curvature and tensor power spectra, following the
conventions of \citet{martin00}.}. We ignore the effects of weak
gravitational lensing.
We consider two models for the
ionization history: no reionization and full reionization at redshift
$z_\rmn{re}=6$. Note that had we adopted a model with earlier reionization,
e.g.\ $z_\rmn{re}\sim 15$ as favoured by \emph{WMAP} data~\citep{kogut03},
the problem of $E$-$B$ mixing described in Section~\ref{windows} would have
been further exacerbated by the additional large-scale $E$ power.

\section{Polarization correlation functions on the sphere}
\label{correlation}

Stokes parameters $Q(\vnhat)$ and $U(\vnhat)$ are defined for a line of sight
$\vnhat$ with the local $x$-axis generated by $\hat{\mathbf{\theta}}$
and the local $y$-axis by $-\hat{\mathbf{\phi}}$. Here $\hat{\mathbf{\theta}}$
and $\hat{\mathbf{\phi}}$ are the basis vectors of a spherical-polar
coordinate system. A right-handed basis
is completed by the addition of the radiation propagation direction
$-\vnhat$. The polarization $P\equiv Q+iU$ is spin $-2$~\citep{newman66}
and can be expanded in spin-$\pm 2$ harmonics as~\citep{seljak97}
\begin{equation}
(Q \pm i U)(\vnhat) = \sum_{lm}(E_{lm} \mp i B_{lm}) {}_{\mp 2}Y_{lm}(\vnhat).
\label{eq:1}
\end{equation}
Reality of $Q$ and $U$ demands $E_{lm}^\ast = (-1)^m E_{l\, -m}$ with
an equivalent result for $B_{lm}$. Under parity transformations,
$(Q\pm i U)(\vnhat) \rightarrow (Q\mp i U)(-\vnhat)$ so that
$E_{lm}$ has parity $(-1)^l$ (electric), but $B_{lm}$ has parity
$(-1)^{l+1}$ (magnetic). The temperature is a scalar field and so
can be expanded in spherical harmonics with multipoles $T_{lm}$.
In an isotropic- and parity-invariant ensemble the
non-vanishing elements of the polarization covariance structure are
\begin{eqnarray}
\langle E_{lm} E_{l'm'}^\ast \rangle &=& \delta_{ll'} \delta_{mm'}
C_l^E, \label{eq:2} \\
\langle B_{lm} B_{l'm'}^\ast \rangle &=& \delta_{ll'} \delta_{mm'}
C_l^B, \label{eq:2b} \\
\langle E_{lm} T_{l'm'}^\ast \rangle &=& \delta_{ll'} \delta_{mm'}
C_l^{TE}. \label{eq:2c}
\end{eqnarray}

If the direction $\vnhat_1$ corresponds to angular coordinates
$(\theta_1,\phi_1)$, and similarly for $\vnhat_2$, then the $\mathrm{SO}(3)$
composition
\begin{equation}
D^{-1}(\phi_1,\theta_1,0)D(\phi_2,\theta_2,0)=D(\alpha,\beta,-\gamma)
\label{eq:3}
\end{equation}
determines $\beta$ ($0 \leq \beta \leq \pi$), the angle between
$\vnhat_1$ and $\vnhat_2$, $\alpha$, the angle required to rotate
$\hat{\bmath{\theta}}(\vnhat_1)$ in a right-handed sense about
$\vnhat_1$ onto the tangent (at $\vnhat_1$) to the geodesic connecting
$\vnhat_1$ and $\vnhat_2$, and $\gamma$, defined in the same manner as
$\alpha$ but at $\vnhat_2$. Making use of the relation
between the Wigner-$D$ matrices (e.g.\ \citealt{varshalovich})
and the spin-weight spherical harmonics,
\begin{equation}
D^l_{-m s} (\phi,\theta,0) = (-1)^m \sqrt{\frac{4\pi}{2l+1}}
{}_s Y_{lm}(\vnhat),
\label{eq:4}
\end{equation}
we obtain the following representation of equation~(\ref{eq:3}):
\begin{equation}
D^l_{ss'}(\alpha,\beta,-\gamma) = \sum_{m} \frac{4\pi}{2l+1}
{}_s Y_{lm}^\ast(\vnhat_1) {}_{s'}Y_{lm}(\vnhat_2).
\label{eq:5}
\end{equation}
With this result, the two-point correlation functions for linear
polarization evaluate to~\citep{ng99}
\begin{eqnarray}
\langle \bar{P}(\vnhat_1) \bar{P}(\vnhat_2) \rangle
&=& \sum_l \frac{2l+1}{4\pi} (C_l^E - C_l^B) d^l_{2\, -2}(\beta), 
\label{eq:6}\\
\langle \bar{P}^\ast(\vnhat_1) \bar{P}(\vnhat_2) \rangle
&=& \sum_l \frac{2l+1}{4\pi} (C_l^E + C_l^B) d^l_{2\, 2}(\beta),
\label{eq:7}\\
\langle T(\vnhat_1) \bar{P}(\vnhat_2) \rangle &=&
\sum_l \frac{2l+1}{4\pi} C_l^{TE} d^l_{2\, 0}(\beta), \label{eq:7b}
\end{eqnarray}
where $d^l_{mn}$ are the reduced $D$-matrices. Note that
$\langle T(\vnhat_1) \bar{P}(\vnhat_2) \rangle =
\langle \bar{P}(\vnhat_1) T(\vnhat_2) \rangle$.
The quantities
\begin{eqnarray}
\bar{P}(\vnhat_1) &\equiv& e^{2i\alpha} P(\vnhat_1), \label{eq:8} \\
\bar{P}(\vnhat_2) &\equiv& e^{2i\gamma} P(\vnhat_2), \label{eq:9}
\end{eqnarray}
are the polarizations defined on local bases with the $x$-direction
along the geodesic between $\vnhat_1$ and $\vnhat_2$. With these rotations,
the correlation functions depend only on the angle $\beta$ between the two
points. Note that $\langle \bar{P}^\ast(\vnhat_1) \bar{P}(\vnhat_2)
\rangle$ is real, which follows from statistical isotropy,
while $\langle \bar{P}(\vnhat_1) \bar{P}(\vnhat_2) \rangle$
and $\langle T(\vnhat_1) \bar{P}(\vnhat_2) \rangle$ are
only real if the universe is parity-invariant in the mean. In the presence
of parity violations,
\begin{eqnarray}
\langle \bar{P}(\vnhat_1) \bar{P}(\vnhat_2) \rangle
&=& \sum_l \frac{2l+1}{4\pi} [(C_l^E - C_l^B -2iC_l^{EB}) \nonumber \\
&& \mbox{} \times d^l_{2\, -2}(\beta)], 
\label{eq:10} \\
\langle T(\vnhat_1) \bar{P}(\vnhat_2) \rangle &=&
\sum_l \frac{2l+1}{4\pi} (C_l^{TE}-iC_l^{TB}) d^l_{2\, 0}(\beta),
\label{eq:10b} 
\end{eqnarray}
where
\begin{equation}
\langle E_{lm} B_{lm}^\ast \rangle = \delta_{ll'}\delta_{mm'}
C_l^{EB}, \quad \langle T_{lm} B_{lm}^\ast \rangle = \delta_{ll'}\delta_{mm'}
C_l^{TB}.
\end{equation}
The correlation functions of the (rotated) Stokes parameters
can be found directly from those for $\bar{P}$. Defining
\begin{eqnarray}
\xi_-(\beta) &\equiv& \langle \bar{P}(\vnhat_1) \bar{P}(\vnhat_2) \rangle,
\label{eq:11} \\
\xi_+(\beta) &\equiv& \langle \bar{P}^\ast(\vnhat_1)\bar{P}(\vnhat_2)\rangle,
\label{eq:12} \\
\xi_X(\beta) &\equiv& \langle T(\vnhat_1) \bar{P}(\vnhat_2) \rangle,
\label{eq:12b}
\end{eqnarray}
we have
\begin{eqnarray}
\langle \bar{Q}(\vnhat_1) \bar{Q}(\vnhat_2) \rangle &=&
\frac{1}{2}[\xi_+(\beta) + \Re \xi_-(\beta)], \label{eq:13} \\
\langle \bar{U}(\vnhat_1) \bar{U}(\vnhat_2) \rangle &=&
\frac{1}{2}[\xi_+(\beta) - \Re \xi_-(\beta)], \label{eq:14} \\
\langle \bar{Q}(\vnhat_1) \bar{U}(\vnhat_2) \rangle &=&
\frac{1}{2} \Im \xi_-(\beta), \label{eq:15} \\
\langle T(\vnhat_1) \bar{Q}(\vnhat_2) \rangle &=&
\Re \xi_X(\beta), \label{eq:15b} \\
\langle T(\vnhat_1) \bar{U}(\vnhat_2) \rangle &=&
\Im \xi_X(\beta). \label{eq:15c}
\end{eqnarray}
%
\begin{figure}
\epsfig{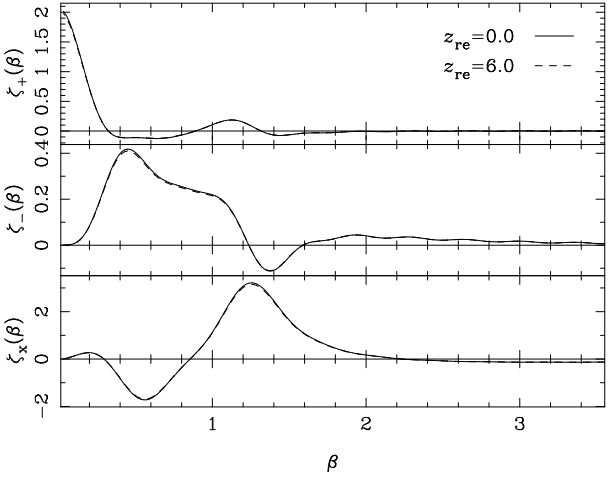}
\caption{The correlation functions
$\xi_+(\beta)$ (top panel), $\xi_-(\beta)$ (middle), and $\xi_X(\beta)$
(bottom). The cosmological model is as described in Sec.~\ref{sec:intro};
the solid lines are for no reionization, while the dashed have complete
reionization with $z_\rmn{re}=6$. The angle $\beta$ is in degrees.
\label{figcpcm}}
\end{figure}
%
In Fig.~\ref{figcpcm} we plot the correlation functions $\xi_\pm(\beta)$
and $\xi_X(\beta)$ for the cosmological models described in
Sec.~\ref{sec:intro}.
The damping of the polarization power on linear scales that are sub-Hubble
at the epoch of reionization (in this case at $z=6$) is just discernible
in the correlation functions in Fig.~\ref{figcpcm}. The additional
large-scale power due to reionization makes a negligible contribution to the
correlation functions for the angular range we have plotted.

We can invert equations~(\ref{eq:7}), (\ref{eq:10}) and (\ref{eq:10b})
using the orthogonality of the $d^l_{mm'}$,
\begin{equation}
\int_{-1}^1 d^l_{mm'}(\beta) d^{l'}_{mm'}(\beta)\, \ud \cos\beta
= \frac{2}{2l+1} \delta_{ll'},
\label{eq:16}
\end{equation}
to determine the power spectra from the correlation functions:
\begin{eqnarray}
C_l^E - C_l^B -2i C_l^{EB} &=& 2\pi \int_{-1}^{1} 
\xi_-(\beta) d^l_{2\, -2}(\beta) \,\ud \cos\beta , \label{eq:17} \\
C_l^E + C_l^B &=&  2\pi \int_{-1}^{1}
\xi_+(\beta) d^l_{2\, 2}(\beta)\, \ud \cos\beta , \label{eq:18} \\
C_l^{TE} + i C_l^{TB} &=& 2\pi \int_{-1}^{1} 
\xi_X(\beta) d^l_{2\, 0}(\beta) \,\ud \cos\beta . \label{eq:18b}
\end{eqnarray}
Since those reduced $D$-matrices that appear in these equations are
are polynomials in $\cos\beta$ for, the integrals can be performed
essentially exactly for band-limited data by Gauss-Legendre quadrature.

\section{Fast correlation function estimators}
\label{estimators}

If we had available unbiased estimates of the various correlation functions
for all angles $0\leq \beta \leq \pi$, we could obtain
unbiased estimates of the power spectra by performing the inversions
in equations~(\ref{eq:17})--(\ref{eq:18b}). Direct evaluation of the
correlation functions~\citep[e.g.][]{sbarra03}
requires $O(N_{\rmn{pix}}^2)$ evaluations and is
complicated by the need to perform a rotation to the appropriate basis
for each pair of points. Here we consider an $O(N_{\rmn{pix}}^{3/2})$ method
based on fast spherical transforms. This method generalises that
of~\citet{szapudi01} to polarization fields.

We consider an arbitrary weighting of the noisy polarization
field $P(\vnhat)$ and the noisy temperature field $T(\vnhat)$
with some weight functions $w_P(\vnhat)$ and $w_T(\vnhat)$ respectively.
The weight is zero for those pixels in regions
that are either not observed or are removed from the map due to foreground
contamination. In this paper we only consider real weighting of the
polarization field, thus preserving the direction of polarization
at any point; relaxing this condition is straightforward if required.
In the presence of instrument noise, the weights allow
for a heuristic pixel-noise weighting of the data. We start with the
following estimators for the signal-plus-noise correlations:
\begin{eqnarray}
\hat{C}_+(\psi) &=& A_P(\psi) \int \ud\vnhat_1 \ud\vnhat_2\,
[\delta(\vnhat_1 \cdot \vnhat_2 - \cos\psi) \nonumber \\
&&\phantom{xxxxxxx} \times w_P(\vnhat_1) w_P(\vnhat_2)
\bar{P}^\ast(\vnhat_1) \bar{P}(\vnhat_2)] , \label{eq:19} \\
\hat{C}_-(\psi) &=& A_P(\psi) \int \ud\vnhat_1 \ud\vnhat_2\,
[\delta(\vnhat_1 \cdot \vnhat_2 - \cos\psi) \nonumber \\
&&\phantom{xxxxxxx}\times w_P(\vnhat_1) w_P(\vnhat_2) \bar{P}(\vnhat_1)
\bar{P}(\vnhat_2)] , \label{eq:20} \\
\hat{C}_X(\psi) &=& A_X(\psi) \int \ud\vnhat_1 \ud\vnhat_2\,
[\delta(\vnhat_1 \cdot \vnhat_2 - \cos\psi) \nonumber \\
&&\phantom{xxxxxxx} \times w_T(\vnhat_1) w_P(\vnhat_2)
T(\vnhat_1) \bar{P}(\vnhat_2)] . \label{eq:20b}
\end{eqnarray}
The delta functions ensure that we only consider those points that
have angular separation $\psi$. The normalisations $A(\psi)$ are chosen
so that our correlation function estimators are unbiased in the absence
of noise. This requires
\begin{eqnarray}
\frac{1}{A_P(\psi)} &=& \int \ud\vnhat_1 \ud\vnhat_2\,
[\delta(\vnhat_1 \cdot \vnhat_2 - \cos\psi) \nonumber \\
&&\mbox{} \times w_P(\vnhat_1) w_P(\vnhat_2)],
\label{eq:21} \\
\frac{1}{A_X(\psi)} &=& \int \ud\vnhat_1 \ud\vnhat_2\,
[\delta(\vnhat_1 \cdot \vnhat_2 - \cos\psi) \nonumber \\
&&\mbox{} \times w_T(\vnhat_1) w_P(\vnhat_2)].
\label{eq:21b}
\end{eqnarray}
These expressions for the correlation functions and normalisation factor
can be simplified by using the completeness relation
\begin{equation}
\sum_{l \geq {\rmn{max}}(|m|,|n|)} \frac{2l+1}{2} d^l_{mn}(\beta)
d^l_{mn}(\psi) = \delta(\cos\beta - \cos\psi)
\label{eq:22}
\end{equation}
to substitute for the delta functions. To evaluate $\hat{C}_+(\psi)$
we set $m=n=2$, so the integrand in equation~(\ref{eq:19}) involves
\begin{eqnarray}
&& w_P(\vnhat_1)\bar{P}^\ast (\vnhat_1) d^l_{22}(\beta) \bar{P}(\vnhat_2)
w_P(\vnhat_2) \nonumber \\
&&\phantom{xxxxxxxxxx} =  \tilde{P}^\ast
(\vnhat_1) D^l_{22}(\alpha,\beta,-\gamma) \tilde{P}(\vnhat_2),
\label{eq:23}
\end{eqnarray}
where $\cos\beta = \vnhat_1\cdot\vnhat_2$, and
we have used equations~(\ref{eq:8}) and (\ref{eq:9}). Here,
$\tilde{P}(\vnhat) \equiv w_P(\vnhat)P(\vnhat)$ is the weighted
polarization field on the (polar-)coordinate basis. We can now use
equation~(\ref{eq:5}) to express the $D$-matrix in terms of spin-weight
harmonics. Performing the angular integrals extracts the spin-weight
2 (pseudo-)multipoles of the weighted, noisy polarization field, defined by
\begin{eqnarray}
\tilde{P}(\vnhat) &=& \sum_{lm}(\tilde{E}_{lm} - i \tilde{B}_{lm})
{}_{-2}Y_{lm}(\vnhat), \label{eq:24} \\
\tilde{P}^\ast(\vnhat) &=& \sum_{lm}(\tilde{E}_{lm} + i \tilde{B}_{lm})
{}_{+2}Y_{lm}(\vnhat), \label{eq:25}
\end{eqnarray}
leaving
\begin{equation}
\hat{C}_+(\psi) = 2\pi A_P(\psi) \sum_{lm}
d^l_{22}(\psi) |\tilde{E}_{lm}  + i \tilde{B}_{lm}|^2.
\label{eq:26}
\end{equation}
Introducing the real pseudo-$C_l$s for the weighted fields:
\begin{eqnarray}
\tilde{C}_l^E &\equiv& \frac{1}{2l+1} \sum_{m} |\tilde{E}_{lm}|^2,
\label{eq:27} \\
\tilde{C}_l^B &\equiv& \frac{1}{2l+1} \sum_{m} |\tilde{B}_{lm}|^2,
\label{eq:28} \\
\tilde{C}_l^{EB} &\equiv& \frac{1}{2l+1} \sum_{m} \tilde{E}_{lm}
\tilde{B}^\ast_{lm} =  \frac{1}{2l+1} \sum_{m} \tilde{B}_{lm}
\tilde{E}^\ast_{lm}, \label{eq:29} \\
\tilde{C}_l^{TB} &\equiv& \frac{1}{2l+1} \sum_{m}\tilde{T}_{lm}
\tilde{B}^\ast_{lm} =  \frac{1}{2l+1} \sum_{m} \tilde{B}_{lm}
\tilde{T}^\ast_{lm},
\label{eq:29b}
\end{eqnarray}
we can write
\begin{equation}
\hat{C}_+(\psi) = 2\pi A_P(\psi) \sum_{l} (2l+1)
d^l_{22}(\psi) (\tilde{C}_l^E + \tilde{C}_l^B).
\label{eq:30}
\end{equation}
To evaluate $\hat{C}_-(\psi)$ we follow the same procedure, but with
$m=-2$ and $n=2$ in equation~(\ref{eq:22}). The result is
\begin{equation}
\hat{C}_-(\psi) = 2\pi A_P(\psi) \sum_l (2l+1)
d^l_{2\, -2}(\psi) (\tilde{C}_l^E - \tilde{C}_l^B -2 i \tilde{C}_l^{EB}).
\label{eq:31}
\end{equation}
Finally, for $\hat{C}_X(\psi)$ we take $m=2$ and $n=0$ to find
\begin{equation}
\hat{C}_X(\psi) = 2\pi A_X(\psi) \sum_l (2l+1)
d^l_{2\, 0}(\psi) (\tilde{C}_l^{TE} - i \tilde{C}_l^{TB}).
\label{eq:31b}
\end{equation}
The normalisation factors $A(\psi)$ can be evaluated by taking
$m=n=0$ in equation~(\ref{eq:22}), e.g.\
\begin{equation}
\frac{1}{A_P(\psi)} = 2\pi \sum_{l\geq 0} (2l+1) P_l(\cos\psi) w_{P,l},
\label{eq:32}
\end{equation}
where
\begin{equation}
w_{P,l} = \frac{1}{2l+1} \sum_m |w_{P,lm}|^2,
\label{eq:33}
\end{equation}
with $w_{P,lm}$ the (spin-0) spherical multipoles of the weight function
$w_P(\vnhat)$. Note that we have used $d^l_{00}(\psi) = P_l(\cos\psi)$
where $P_l(x)$ is a Legendre polynomial. Once the mean noise
contribution (noise bias) is removed
from the estimators $\hat{C}(\psi)$
[leaving unbiased estimators of the signal
correlation functions $\xi(\psi)$], we can use
equations~(\ref{eq:17})--(\ref{eq:18b})
to compute estimates of the power spectra. The real parts of
$\hat{\xi}(\psi)$ give estimates of $C_l^E$,
$C_l^B$ and $C_l^X$, while the imaginary parts of $\hat{\xi}_-(\psi)$
and $\hat{\xi}_X(\psi)$ can be used to
estimate $C_l^{EB}$ and $C_l^{TB}$ and hence test for parity violations.

The full set of pseudo-multipoles can be obtained efficiently in
$O(N_{\rmn{pix}}^{3/2} \log N_{\rmn{pix}})$ operations using fast spherical
transforms such as those implemented in the {\sevensize HEALP}ix\footnote{%
http://www.eso.org/science/healpix/} and
{\sevensize IGLOO}\citep{igloo} packages. (Our current implementation employs
{\sevensize HEALP}ix.) To remove the noise bias from $\hat{C}(\psi)$
it is generally most efficient to resort to Monte-Carlo simulations of
pure noise fields~\citep{szapudi01}. (An exception is the case where the noise
is uncorrelated between pixels; see below for details.) The ensemble mean
of these pure-noise correlation functions can be subtracted from
$\hat{C}(\psi)$ to yield (asymptotically) unbiased estimates of the
signal correlation functions. Monte-Carlo estimation of the noise bias
provides a robust means of dealing with discretisation effects due
to the chosen pixelisation. Monte-Carlo methods also offer the simplest method
of computing the variance of the power spectrum estimates. In the
presence of uncorrelated noise it is straightforward to proceed analytically
with the noise contribution to the variance, but the cosmic variance
contribution is complicated by the presence of signal correlations.

For the simple case
of noise that is uncorrelated between pixels it is straightforward to compute
the noise bias analytically. For simplicity consider noise that is uncorrelated
between $Q$, $U$ and $T$, and has equal variance in $Q$ and $U$.
If the noise variance of the Stokes parameters per solid angle
is $\sigma_P^2(\vnhat_p)$,
then in the continuum limit the polarization noise correlations can be
summarised by
\begin{eqnarray}
\langle P_N(\vnhat_1) P_N^\ast(\vnhat_2) \rangle &=& 2 \sigma_P^2
\delta(\vnhat_1 - \vnhat_2) , \label{eq:34} \\
\langle P_N(\vnhat_1) P_N(\vnhat_2) \rangle &=& 0, \label{eq:35} \\
\langle T_N(\vnhat_1) P_N(\vnhat_2) \rangle &=& 0, \label{eq:35b} 
\end{eqnarray}
where $P_N(\vnhat)$ is the spin $-2$ noise,
and $T_N(\vnhat)$ is the noise on the temperature.
As the noise is uncorrelated between pixels its mean effect on correlation
function estimates is confined to zero separation:
\begin{eqnarray}
\langle \Delta \hat{C}_+(\psi)\rangle &=& A_P(0) \delta(1-\cos\psi)
\nonumber \\
&&\mbox{} \times
\int\ud\vnhat \, w_P^2(\vnhat) 2 \sigma_P^2(\vnhat), \label{eq:36} \\
\langle \Delta \hat{C}_-(\psi)\rangle &=& 0, \label{eq:37} \\
\langle \Delta \hat{C}_X(\psi)\rangle &=& 0, \label{eq:37b}
\end{eqnarray}
with
\begin{equation}
\frac{1}{A_P(0)} = 2\pi \int \ud \vnhat \, w_P^2(\vnhat).
\label{eq:38}
\end{equation}
Here, $\langle \Delta \hat{C}\rangle$ is the mean noise contribution
to the estimators $\hat{C}$.
Making use of $d^l_{mm'}(0) = \delta_{mm'}$ we find that the non-zero
noise biases in the estimates of the power spectra in the continuum limit are
\begin{equation}
\langle \Delta \hat{C}_l^E \rangle =\langle \Delta \hat{C}_l^B \rangle
= \frac{\int \ud\vnhat\, w_P^2(\vnhat) \sigma_P^2(\vnhat)}{%
\int \ud\vnhat\, w_P^2(\vnhat)}. 
\label{eq:39}
\end{equation}
However, we would recommend removing the noise bias with Monte-Carlo
techniques even for simple, uncorrelated noise. This is to ensure that
the effective band limit introduced on the noise by computing
the correlation functions via pseudo-$C_l$s up to some finite $l_{\rmn{max}}$
is properly accounted for.

\subsection{Application to large-area surveys}
\label{fullsky}

As an application of our method we consider extracting the power spectra
from simulated maps obtained with a full-sky survey with pixel noise
similar to that expected for \emph{Planck}.
To be specific, we assumed uncorrelated
pixel noise on $Q$ and $U$ with r.m.s.\ $6.95\,\mu\rmn{K}$ in a 10-arcmin
by 10-arcmin pixel. We adopted a beam size of 10 arcmin, somewhat larger than
the polarization-sensitive channels of the \emph{Planck}
High-Frequency Instrument,
so to ensure oversampling of the beam at {\sevensize HEALP}ix resolution
$N_\rmn{side}=1024$. We ignored the variation in pixel noise across the
map, but this could easily be included in our simulations at no additional
computational cost. Noise correlations could also be included easily if
fast simulation of noise realisations were possible. We made
a constant-latitude Galactic cut of $\pm 20\degr$. The underlying
cosmological model was as described in Sec.~\ref{sec:intro} and we assumed
reionization at $z=6$. We adopted a uniform weighting scheme
motivated by the constant variance of the noise.

\begin{figure}
\epsfig{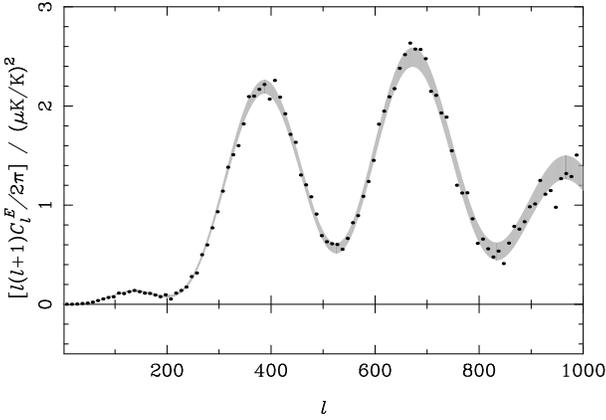}
\caption{Recovered $C_l^E$ power spectrum for a large-area survey with
a $\pm 20\degr$ Galactic cut. The points are flat band-power estimates,
with $\Delta l= 10$,
from a single simulation; the shaded region shows the $\pm \sigma$ region
on the basis of equation~(\ref{eq:error}).
\label{figlargeareaE}}
\end{figure}

The recovered power spectrum $C_l^E$ is shown in Fig.~\ref{figlargeareaE}.
We computed estimates of the correlation function at the roots of a Legendre
polynomial from the pseudo-$C_l$s (obtained with the fast spherical transforms
in {\sevensize HEALP}ix). Inversion of the correlation functions was performed
with Gauss-Legendre integration. We averaged the recovered $l(l+1)C_l$ to
form flat band-power estimates with a $\Delta l = 10$, and the results of
one simulation are shown as the points in Fig.~\ref{figlargeareaE}.
We adopted a $\Delta l= 10$ to ensure the errors were essentially
uncorrelated.
The shaded area in Fig.~\ref{figlargeareaE}, centred on the true spectrum
smoothed with a 10-arcmin beam, 
encloses the $\pm \sigma$ error region based on the
rule of thumb (generalised from~\citealt{hivon02} for the temperature case)
\begin{equation}
\Delta C_l^E \approx  \sqrt{\frac{2}{\nu_l}} (C_l^E + N_l).
\label{eq:error}
\end{equation}
Here $N_l$ is the (full-sky) noise power spectrum, and $\nu_l \equiv \Delta l
(2l+1)f_\rmn{sky}w_2^2/w_4$ is the 
effective number of degrees of freedom in a band of width $\Delta l$ on a
fraction of the sky $f_\rmn{sky}$, where
$4\pi w_i f_\rmn{sky} \equiv \int w_P^i(\vnhat) \, \ud \vnhat$. Note that
$\nu_l$ takes no account of the loss of degrees of freedom associated with
disentangling $E$ and $B$ polarization, since the fractional loss of modes is
small for $f_\rmn{sky}$ close to unity~\citep{lewis02}. The scatter of
points in the simulation is broadly consistent with that expected on the
basis of the theoretical errors. A more detailed analysis of the
optimality of our method must await comparison with optimal,
maximum-likelihood codes when these become available at sufficiently
high resolution.

\section{Window functions from incomplete correlation functions}
\label{windows}

In this section we construct the window functions that arise when the
correlation functions can only be estimated over a limited angular range.
We shall concentrate on the polarization auto-correlations; the
generalisation to the polarization-temperature cross-correlation is
straightforward.

If unbiased estimates of the correlation functions are available over the
full angular range $(0,\pi)$, they can easily be inverted to obtain
unbiased power spectra. This case would describe full-sky experiments with
a cut excising less than a $90^\circ$ band about the Galactic plane.
In the case where the correlation functions cannot be estimated for all
separations $\psi$, estimating the power spectra by direct
integration (e.g.\ equations~\ref{eq:17} and \ref{eq:18}) over the
observed range will introduce window functions ${}_{\pm 2}K_{ll'}$
such that
\begin{equation}
\langle \hat{C}_l^E \pm \hat{C}_l^B \rangle = \sum_{l'} {}_{\pm 2}K_{ll'}
(C_{l'}^E \pm C_{l'}^B).
\label{eq:40}
\end{equation}
Fourier ringing can be reduced by pre-multiplying the correlation
functions with a scalar apodizing function $f(\psi)$ prior to integration,
in which case the window functions take the form
\begin{equation}
{}_{\pm 2}K_{ll'} \equiv \frac{2l'+1}{2}\int f(\beta) d^l_{2\, \pm 2}(\beta)
d^{l'}_{2\, \pm 2}(\beta) \, \ud\cos\beta,
\label{eq:41}
\end{equation}
where the integral is over the range of angles for which the correlation
functions can be estimated. Note that the window functions are not
symmetric but rather satisfy
\begin{equation}
(2l+1) {}_{\pm 2}K_{ll'} = (2l'+1){}_{\pm 2}K_{l'l}.
\end{equation}
Introducing the sum and difference window
functions, ${}_\pm K_{ll'} \equiv ({}_{2}K_{ll'} \pm {}_{-2}K_{ll'})/2$,
the means of the estimated power spectra are related to the
true spectra by
\begin{eqnarray}
\langle \hat{C}_l^E \rangle &=& \sum_{l'} ({}_+ K_{ll'} C_{l'}^E +
{}_- K_{ll'} C_{l'}^B) , \label{eq:42} \\
\langle \hat{C}_l^B \rangle &=&
 \sum_{l'} ({}_- K_{ll'} C_{l'}^E +
{}_+ K_{ll'} C_{l'}^B) . \label{eq:43}
\end{eqnarray}
The window function ${}_- K_{ll'}$ controls the mixing of $E$ and $B$
polarization. Recent results from DASI~\citep{kovac02} are in line with
theoretical expectations that $B$ polarization should be sub-dominant, so that
cross contamination due to partial sky effects is proportionately more
troubling for $B$ polarization than for $E$. While not presenting a
fundamental problem for cosmological parameter extraction, a non-zero
${}_- K_{ll'}$ makes interpretation (and presentation) of the estimated
$C_l^B$ awkward. Mixing can obviously be eliminated by pre-multiplying
the estimates $\hat{C}_l^E \pm \hat{C}_l^B$ with the inverse of the window
functions ${}_{\pm 2}K_{ll'}$, but this inversion is awkward in practice
due to the ill-conditioned nature of the window functions when
coverage of the correlation functions is incomplete. In
Section~\ref{sec:eb} we introduce a simple, robust technique
for extracting the power spectra from correlation functions which eliminates
mixing in the mean (i.e.\ produces a zero ${}_- K_{ll'}$). Before
turning to that, in the following subsections we first explore the properties
of the window functions given by equation~(\ref{eq:41}), and the circumstances
under which mixing is significant.

\subsection{General properties}

If we define the apodizing function $f(\psi)$ to be zero outside the
observed range of $\psi$, and perform a Legendre expansion
\begin{equation}
f(\psi) = \sum_{l\geq 0} \frac{2l+1}{2} f_l P_l(\cos\psi),
\label{eq:44}
\end{equation}
the window function ${}_2 K_{ll'}$ reduces to
\begin{equation}
{}_2 K_{ll'} = \frac{2l'+1}{2} \sum_L (2L+1)f_L
\left( \begin{array}{ccc} l & l' & L \\ 2 & -2 & 0 \end{array}\right)^2 ,
\label{eq:45}
\end{equation}
while ${}_{-2}K_{ll'}$ has an additional factor of $(-1)^{(l+l'+L)}$ in the
summation. The array in brackets in equation~(\ref{eq:45}) is a
Wigner 3-$j$ symbol arising from the integral of a product of three
rotation matrices. If the apodizing function is effectively band limited to
$L_\rmn{max}$, the window functions vanish for $|l-l'| > L_\rmn{max}$.

The normalisation $\sum_{l'} {}_{\pm 2}K_{ll'}$
of the window functions is also of some interest.
For ${}_2 K_{ll'}$ we can perform the sum over $l'$ in
equation~(\ref{eq:45}) directly using the orthogonality of the
3-$j$ symbols~\citep{varshalovich} to find
\begin{equation}
\sum_{l'} {}_2 K_{ll'} = \sum_{L \geq 0} \frac{2L+1}{2} f_L =
f(0).
\label{eq:46}
\end{equation}
The last equality follows from $P_l(1)=1$, or, more directly, by employing
$\sum_{l'} (l'+1/2) d^{l'}_{22}(\beta) = \delta(\cos\beta-1)$ in
equation~(\ref{eq:41}). The normalisation of ${}_{-2}K_{ll'}$ is a little
more involved. We start with the result
\begin{equation}
\sum_{l'} \frac{2l'+1}{2} d^{l'}_{2\, -2}(\beta) = \delta(\cos\beta-1)
+ \csc^2(\beta/2),
\label{eq:47}
\end{equation}
which follows by summing equation~(\ref{eq:73}) of Section~\ref{sec:eb} over
$l'$. If we now sum equation~(\ref{eq:41}) over $l'$, and use
equation~(\ref{eq:47}), we find that
\begin{equation}
\sum_{l'} {}_{-2} K_{ll'} = \int f(\beta) \csc^2(\beta/2)d^l_{2\, -2}(\beta)
\, \ud \cos\beta,
\label{eq:48}
\end{equation}
where we have used $d^l_{2\, -2}(0)=0$ [and the assumed regularity of
$f(\beta)$]. The function $\csc^2(\beta/2)d^{l}_{2\, -2}(\beta)$
is a polynomial in
$\cos\beta$ and so the integral can easily be evaluated numerically by e.g.\
Gauss-Legendre integration for smooth apodizing functions. To make
further progress analytically we insert the Legendre expansion of
$f(\beta)$ in equation~(\ref{eq:48}) and use the differential representation
of the reduced $D$-matrices (e.g.\ Section 4.3.2 of \citealt{varshalovich}).
Repeated integration by parts then establishes the result
\begin{eqnarray}
\sum_{l'} {}_{-2}K_{ll'} &=& \sum_{L\leq l} \frac{2L+1}{2}f_L\left(
1-4\frac{L(L+1)}{l(l+1)} \right. \nonumber \\
&&\mbox{} \left. + 3 \frac{(L+2)!}{(L-2)!}\frac{(l-2)!}{(l+2)!}
\right).
\label{eq:49} 
\end{eqnarray}
If $f(\beta)$ is effectively band-limited, for $l \gg L_\rmn{max}$ we have
$\sum_{l'} {}_{-2} K_{ll'} \approx f(0)$. In this limit, the normalisation
of ${}_- K_{ll'}$ is much smaller than that of ${}_+ K_{ll'}$, and mixing of
$E$ and $B$-power is suppressed in the mean~\citep{bunn02}.

\subsection{No apodization}

Consider the case where the correlation functions can be estimated in the
range $(0,\beta_\rmn{max})$. If we apply no apodization to the correlation
functions, we obtain window functions
\begin{equation}
{}_{\pm 2} W_{ll'} \equiv \frac{2l'+1}{2} \int_{\cos\beta_\rmn{max}}^1
d^l_{2\, \pm 2}(\beta) d^{l'}_{2\, \pm 2}(\beta) \, \ud\cos\beta.
\label{eq:44-1}
\end{equation}
For $l\neq l'$ this integral can be evaluated directly since the
$d^l_{mn}$ are eigenfunctions of a self-adjoint operator. The result
is
\begin{eqnarray}
{}_{\pm 2} W_{ll'} &=& \frac{2l'+1}{2} \frac{\cos^2\beta_\rmn{max}}{%
l(l+1)-l'(l'+1)} \left(\frac{\ud\, d^l_{2\, \pm 2}}{\ud\cos\beta}
d^{l'}_{2\, \pm 2} \right. \nonumber \\
&&\mbox{} \left.\left.
- \frac{\ud\, d^{l'}_{2\, \pm 2}}{\ud\cos\beta} d^l_{2\,\pm 2}\right)
\right|_{\beta_\rmn{max}}, \quad l\neq l'.
\label{eq:45-1}
\end{eqnarray}
For $l=l'$ the integral can be evaluated recursively as described in
Appendix C of~\citet{lewis02}. The window function
${}_- W_{ll'} \equiv ({}_{2} W_{ll'} - {}_{-2}W_{ll'})/2$ can be evaluated
directly for all $l$ and $l'$~\citep{lewis02}:
\begin{equation}
{}_- W_{ll'} = \frac{2l'+1}{2}(u_l u_{l'} + v_l v_{l'})|_{\beta_\rmn{max}},
\label{eq:46-1}
\end{equation}
where the vectors
\begin{eqnarray}
u_l(\beta) &\equiv& \sqrt{\frac{(l-2)!}{(l+2)!}} \sin\beta \frac{\ud}{\ud\beta}
\left(\frac{d^l_{20}}{\sin\beta}\right), \label{eq:47-1} \\
v_l(\beta) &\equiv& \sqrt{\frac{(l-2)!}{(l+2)!}} \frac{\sqrt{3}}{\sin\beta}
d^l_{20}(\beta). \label{eq:48-1}
\end{eqnarray}
Both vectors vanish for $\beta_\rmn{max}=\pi$ to ensure that
${}_- W_{ll'}=0$ when the full angular range $(0,\pi)$ is considered.

\begin{figure*}
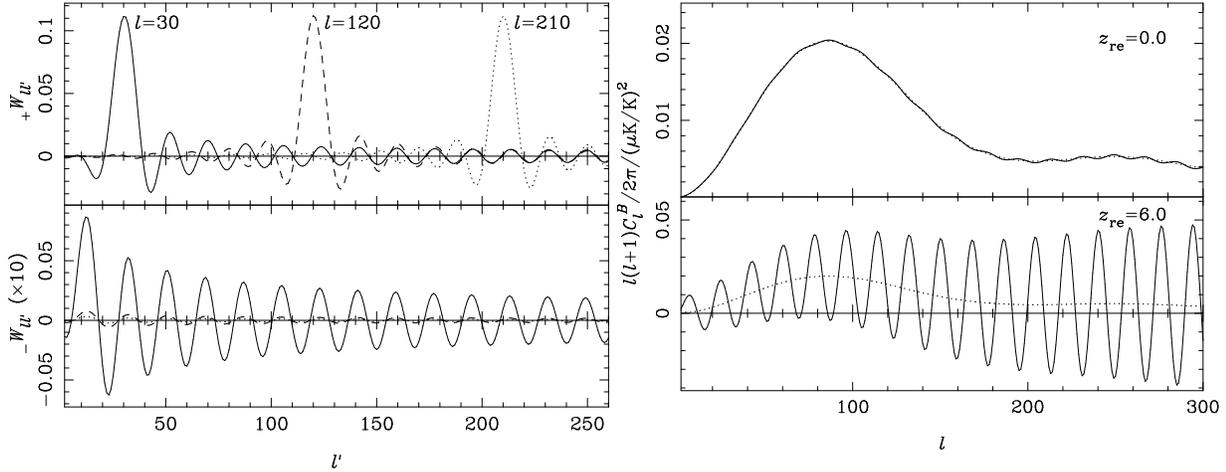

\epsfig{figure=windows_noapodise.ps,angle=-90,width=8cm}
\epsfig{figure=rec_b_noapodise.ps,angle=-90,width=8cm}
\caption{Left: representative rows of the window functions
${}_{+}W_{ll'}$ (top panel) and ${}_- W_{ll'}$ (bottom panel)
when the correlation functions are known in the angular
range $(0,20^\circ)$, and no apodization is applied. The solid lines are
for $l=30$, the dashed for $l=120$ and the dotted for $l=210$.
Right: mean recovered $C_l^B$ (solid lines), obtained from
the convolution in equation~(\ref{eq:43}) with the windows
${}_{\pm}W_{ll'}$, compared to the true $C_l^B$ (dashed lines).
The top panel has no reionization while the bottom panel is a model
with full reionization at $z=6$.
\label{fig1}}
\end{figure*}

Some representative rows of the window functions ${}_\pm W_{ll'}$ are shown
in Fig.~\ref{fig1} for $\beta_\rmn{max} = 20^\circ$ (corresponding
to e.g.\ observations over a circular patch of radius $10^\circ$). Note
that ${}_+W_{ll'}$ is localised around $l=l'$ (with width varying inversely
with $\beta_\rmn{max}$), while ${}_- W_{ll'}$ shows no localisation. 
Equation~(\ref{eq:46-1}) shows that, considered as a matrix,
${}_- W_{ll'}$ is of rank 2, so the rows of the window function are
constructed from linear combinations of $u_{l'}$ and $v_{l'}$. The
approximate scaling of ${}_-W_{ll'}$ with $l$ for fixed $l'$, that is
evident in Fig.~\ref{fig1}, arises because the vector $u_l$ oscillates
with larger amplitude than $v_l$ for $l \ga 1/\beta_\rmn{max}$.

In Fig.~\ref{fig1} we also show the mean of the estimated power spectrum
$\langle \hat{C}_l^B \rangle$ obtained by multiplying the window functions
${}_{\pm} W_{ll'}$ with the true $C_l^E$ and $C_l^B$ (equation~\ref{eq:43})
for the cosmological models detailed in Sec.~\ref{sec:intro}. The true
spectra are convolved with a Gaussian beam of full-width 10 arcmin
at half-maximum. In the case of no reionization, the mean of the
recovered $C_l^B$ is a faithful representation of the true spectra.
This is because (i) the inner products between $C_{l'}^E$ and either
of $(l'+1/2)u_{l'}$
and $(l'+1/2)v_{l'}$ are sufficiently small that the leakage from
$E$ polarization causes only a small amplitude oscillation in the recovered
$C_l^B$; and (ii) for ${}_+ W_{ll'}$ sufficiently localised compared to the
scale of features in $C_{l'}^B$, we can approximate their product by
\begin{equation}
\sum_{l'} {}_+ W_{ll'} C_{l'}^B \approx C_l^B \sum_{l'} {}_+ W_{ll'}
\approx C_l^B\quad (l \gg 1/\beta_\rmn{max}). 
\label{eq:49-1}
\end{equation}
For the second approximation note that the window function ${}_+ W_{ll'}$
inherits its normalisation from that
of ${}_{2} W_{ll'}$ (which is unity) and ${}_{-2}W_{ll'}$. For the
latter we use equation~(\ref{eq:48}) and the differential representation
of $d^l_{2\, -2}$ to find
\begin{eqnarray}
&&\sum_{l'} {}_{-2} W_{ll'} = 1 + 2\sqrt{\frac{(l-2)!}{(l+2)!}}
\cot(\beta_\rmn{max}/2) \nonumber \\
&&\mbox{} \times\left[\sqrt{l(l+1)} d^l_{1\, -2}(\beta)
- \cot(\beta/2) d^l_{0\, -2}(\beta)\right]_{\beta_\rmn{max}}. 
\label{eq:50}
\end{eqnarray}
For large $l \gg 1/\beta_\rmn{max}$ we find $\sum_{l'}{}_{-2}W_{ll'} \approx
1$ (see also the discussion after equation~\ref{eq:49}). Reionized models are
more problematic since they have additional large-scale power in $E$
polarization (and so are more sensitive to the truncation of the
correlation functions at $\beta_\rmn{max}$). The effect of this large-scale
power can clearly be seen in Fig.~\ref{fig1} for the model with reionization
at $z=6$. The large amplitude oscillations in the recovered $C_l^B$
trace those of the vector $u_l(\beta_\rmn{max})$ at large $l$.

\subsection{Gaussian apodization}

The Fourier ringing evident in the window functions in Fig.~\ref{fig1}
can be reduced by apodizing the correlation functions. Here we consider
Gaussian apodizing functions, i.e.\
\begin{equation}
f(\beta) = e^{-\beta^2/(2\sigma^2)}.
\label{eq:51}
\end{equation}
The half-width at half-maximum is $\sigma\sqrt{2\ln 2}$ which should be small
compared to the cut off $\beta_\rmn{max}$ in the correlation functions for
effective apodizing. The window functions, accounting for apodization and the
finite range $(0,\beta_\rmn{max})$ of the observed correlation functions,
can be written as matrix products:
\begin{equation}
{}_{\pm 2}K_{ll'} = \sum_{L} {}_{\pm 2}F_{lL} \, {}_{\pm 2}W_{Ll'},
\label{eq:52}
\end{equation}
where
\begin{equation}
{}_{\pm 2}F_{ll'} \equiv \frac{2l'+1}{2}\int_{-1}^1 f(\beta) d^l_{2\, \pm 2}
(\beta) d^{l'}_{2\, \pm 2}(\beta) \, \ud\cos\beta.
\label{eq:53}
\end{equation}
Note that the full window functions ${}_{\pm 2}K_{ll'}$ are insensitive
to the behaviour of the apodizing function for $\beta > \beta_\rmn{max}$.
Note also that the order of the matrix product in equation~(\ref{eq:52})
is irrelevant since the window functions commute. If the apodizing function is
narrow compared to $\beta_\rmn{max}$ we expect
${}_{\pm 2}K_{ll'} \approx {}_{\pm 2}F_{ll'}$.

\begin{figure*}
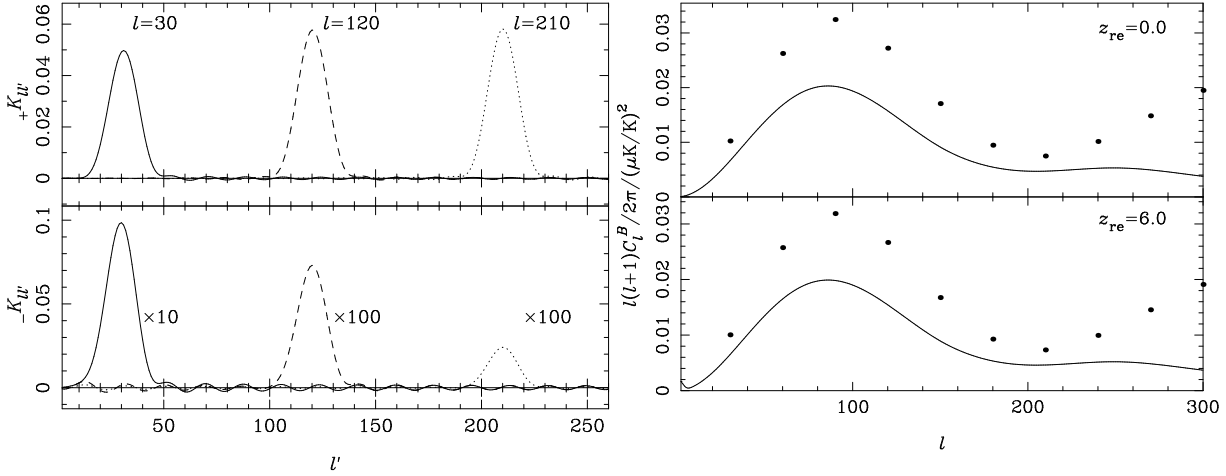

\epsfig{figure=windows_apodise.ps,angle=-90,width=8cm}
\epsfig{figure=rec_b_apodise.ps,angle=-90,width=8cm}
\caption{Left: representative rows of ${}_{\pm} K_{ll'}$ for
$\beta_\rmn{max}=20^\circ$ and Gaussian apodization with half-width at
half-maximum equal to $\beta_\rmn{max}/2$. Right: mean recovered
$C_l^B$ (points) compared to the true $C_l^B$ (lines) with (bottom)
and without (top) reionization.
\label{fig2}}
\end{figure*}

In Fig.~\ref{fig2} we show representative rows of the sum and difference
window functions,
${}_\pm K_{ll'}$, for $\beta_\rmn{max}=20^\circ$ and a Gaussian apodizing
function with half-width at half-maximum equal to $\beta_\rmn{max}/2$ [so
$f(\beta_\rmn{max})=1/16$]. These are well approximated by Gaussians
centred on $l=l'$ with width $1/\sigma$. The amplitude of the difference
window functions are much smaller than those of the sum, and for large
$l$ the ratio of amplitudes $\propto 1/ (l\sigma)^2$.

To understand this behaviour, consider the limit where $\sigma \ll 1$
and $l\sigma \gg 1$. In this flat-sky limit we can
approximate the reduced $D$-matrices by Bessel functions~\citep{varshalovich}:
\begin{equation}
d^l_{22}(\beta) \approx J_0(l\beta), \quad
d^l_{2\, -2}(\beta) \approx J_4(l\beta),
\label{eq:54}
\end{equation}
and the window functions ${}_{\pm 2}F_{ll'}$ are easily computed to be
\begin{eqnarray}
{}_2 F_{ll'} &\approx& l' \sigma^2
e^{-\sigma^2 (l^2+l'{}^2)/2} I_0(ll'\sigma^2),
\label{eq:55} \\
{}_{-2} F_{ll'} &\approx& l' \sigma^2
e^{-\sigma^2 (l^2+l'{}^2)/2} I_4(ll'\sigma^2).
\label{eq:56}
\end{eqnarray}
The leading-order asymptotic expansions of ${}_{\pm}F_{ll'}$ ($ll'\sigma^2
\gg 1$) then follow:
\begin{eqnarray}
{}_+ F_{ll'} &\sim& \sqrt{\frac{l'\sigma^2}{2\pi l}} e^{-(l-l')^2\sigma^2/2},
\label{eq:57} \\
{}_- F_{ll'} &\sim& \frac{4}{ll'\sigma^2} \sqrt{\frac{l'\sigma^2}{2\pi l}}
e^{-(l-l')^2\sigma^2/2},
\label{eq:58}
\end{eqnarray}
which reproduce the behaviour seen in Fig.~\ref{fig2}. The window function
${}_2 F_{ll'}$ is normalised to unity by virtue of equation~(\ref{eq:46}).
The normalisation of ${}_{-2} F_{ll'}$ can be calculated for all $l$
in terms of modified spherical Bessel functions by approximating
$f(\beta) \approx \exp[-(1-\cos\beta)/\sigma^2]$ in equation~(\ref{eq:48}).
However, the result is cumbersome so we shall only give its asymptotic form
here (valid for $\sigma \ll 1$ and $l\sigma \gg 1$):
\begin{eqnarray}
\sum_{l'} {}_{-2}F_{ll'} &=& 1- \frac{4}{l^2\sigma^2}\left(2 +
e^{-l^2\sigma^2/2}\right) \nonumber \\
&&\mbox{} + \frac{24}{l^4\sigma^4}\left(1 -
e^{-l^2\sigma^2/2}\right).
\label{eq:59}
\end{eqnarray}
(This result also follows directly from integrating equation~\ref{eq:56}
over $l'$.) Asymptotically, we then have $\sum_{l'} {}_- F_{ll'} \sim
4 /(l\sigma)^2$, consistent with Fig.~\ref{fig1}.

We also show in Fig.~\ref{fig2} the mean of the recovered $C_l^B$ for
a range of $l$ values with
Gaussian apodizing (corresponding to the window functions in the left-hand
panels of Fig.~\ref{fig2}). The spacing of points is chosen to reflect the
$l$-range over which recovered power spectra should be roughly decorrelated.
While apodizing has clearly removed the high-frequency oscillations in the
mean of the recovered $C_l^B$, it has done so at the expense of introducing
considerable bias due to leakage from $E$ polarization. As expected,
apodization has reduced the sensitivity of the recovered $C_l^B$ to the level
of large-scale power in $E$ polarization
(and hence reionization; c.f.\ Fig.~\ref{fig1}), but has replaced it with
a local bias $\approx 4 C_l^E /(l\sigma)^2$. The bias becomes non-local
in models with sufficiently early reionization, where the level of large-scale
$E$ power can be such that it is transmitted to the recovered $B$ power
spectrum for all $l$ through the low-$l$ tail of the window function.

\begin{figure*}
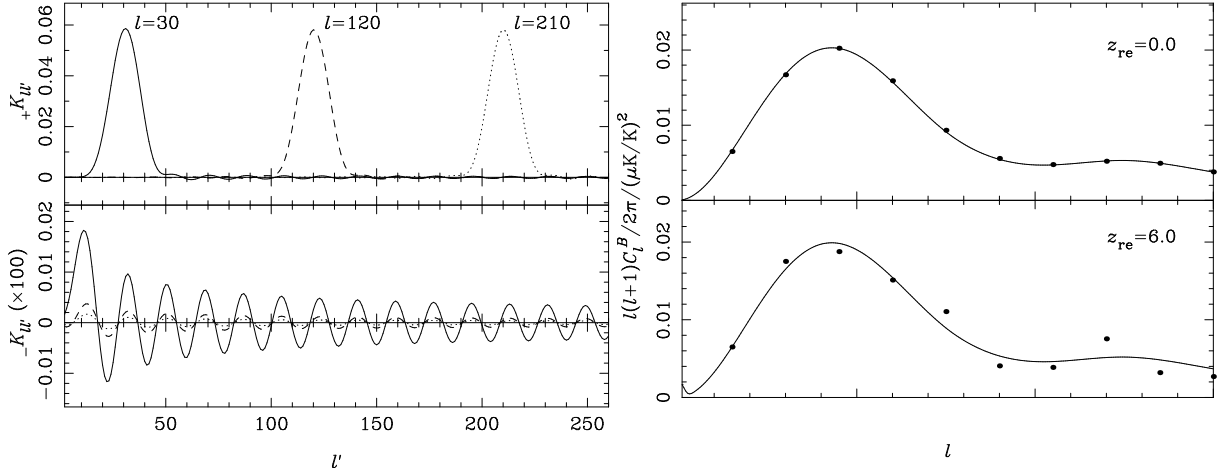

\epsfig{figure=windows_postapodise.ps,angle=-90,width=8cm}
\epsfig{figure=rec_b_postapodise.ps,angle=-90,width=8cm}
\caption{Window functions (left) and mean of recovered $C_l^B$ (right)
obtained by post-convolving the results in Fig.~\ref{fig1} with the
Gaussian asymptotic approximation to ${}_+ F_{ll'}$.
\label{fig3}}
\end{figure*}

A more effective way to reduce oscillations in $C_l^B$ without
introducing additional bias due to $E$-$B$ leakage is to recover the power
spectra with no apodization, and then to post-convolve with a suitably-wide
smoothing function. Such an approach is illustrated in Fig.~\ref{fig3},
where we have post-convolved the results in Fig.~\ref{fig1} with the
asymptotic form of ${}_+ F_{ll'}$ (equation~\ref{eq:57}). The Gaussian
smoothing produces well-localised ${}_+ K_{ll'}$ window functions, removing
the Fourier ringing from ${}_+ W_{ll'}$, and reduces the amplitude of
${}_- W_{ll'}$ by a factor $\sim 40$. Only a low-frequency, oscillatory bias
remains in the recovered $C_l^B$ in the model with reionization. Whether this
bias is significant depends on the level of reionization (and
primordial $B$ polarization).

\section{Removing $E$-$B$ leakage}
\label{sec:eb}

Although we are able to reduce the level of cross-contamination in the
recovered power spectra by simply post-convolving them with a suitably-wide
smoothing function, it is actually straightforward to remove this $E$-$B$
mixing exactly in the mean. \cite{crittenden02} showed how to construct
correlation functions on small patches of the sky that contain only
$E$ or only $B$ modes in the mean. (Their work was in the context of
weak gravitational lensing, but their results are equally applicable to
CMB polarization.) In this section we extend the central result
of~\citeauthor{crittenden02} to the sphere, so we are able to handle
large-angle polarization signals, and demonstrate the methods with
simulations for an experiment similar to BICEP.

We begin by considering the function
\begin{equation}
\xi(\beta) \equiv \sum_l \frac{2l+1}{4\pi} (C_l^E + C_l^B) d^l_{2\, -2}(\beta).
\label{eq:67}
\end{equation}
If we had access to $\xi(\beta)$ over some range of scales we could
combine with the real part of $\xi_-(\beta)$ to extract the function
\begin{equation}
\frac{1}{2}[\xi(\beta)-\Re \xi_-(\beta)] = \sum_{l}\frac{2l+1}{4\pi}
C_l^B d^l_{2\, -2}(\beta),
\label{eq:68}
\end{equation}
which depends only on $B$-polarization. We could thus recover an
estimate of $C_l^B$ by integrating unbiased estimates of $\xi(\beta)-
\xi_-(\beta)$ against $d^l_{2\, -2}$ [with appropriate apodization $f(\beta)$]:
\begin{equation}
\hat{C}_l^B = 2\pi \int \frac{1}{2}[\hat{\xi}(\beta)-\Re\hat{\xi}_-(\beta)]
f(\beta) d^l_{2\, -2}(\beta)\, \ud\cos\beta.
\label{eq:69}
\end{equation}
Such an estimate would contain no contamination from $E$ polarization
in the mean (i.e.\ the difference window function ${}_-K_{ll'}$ would
vanish for arbitrary apodization). A similarly unbiased estimate of
$C_l^E$ could be obtained by considering $\xi(\beta)+\Re \xi_-(\beta)$.

The result we now prove is that the function $\xi(\beta)$ can be
obtained in the range $(0,\beta_{\rmn{max}})$ from the correlation function
$\xi_+(\beta)$ in the same range by quadrature. We start by
inserting equation~(\ref{eq:18}) into the summand of
equation~(\ref{eq:67}) which gives
\begin{equation}
\xi(\beta) = \int_{-1}^{1} \ud\cos\beta' \, \xi_+(\beta') \sum_{l}
\frac{2l+1}{2} d^l_{2\, -2}(\beta) d^l_{22}(\beta').
\label{eq:70}
\end{equation}
Our strategy for simplifying the summation in this equation is to express
$d^l_{2\, -2}(\beta)$ in terms of integrals involving $d^l_{2\, 2}(\beta)$,
and then perform the summation with the completeness relation,
equation~(\ref{eq:22}). Making repeated use of the recursion
relation~\citep{varshalovich}
\begin{eqnarray}
\lefteqn{\frac{-m+m'\cos\beta}{\sin\beta} d^l_{mm'}(\beta) =
\frac{1}{2}\sqrt{(l+m')(l-m'+1)} d^l_{m m'-1}(\beta)} \nonumber \\
&& + \frac{1}{2}\sqrt{(l-m')(l+m'+1)} d^l_{m m'+1}(\beta)
\label{eq:71}
\end{eqnarray}
and the relation
\begin{eqnarray}
\lefteqn{\frac{\ud}{\ud\beta}d^l_{mm'}(\beta) + \frac{m-m'\cos\beta}{\sin\beta}
d^l_{mm'}(\beta)} \nonumber \\
&&\phantom{xxxx} = -\sqrt{(l-m')(l+m'+1)} d^l_{m m'+1}(\beta),
\label{eq:72}
\end{eqnarray}
we find that
\begin{eqnarray}
\lefteqn{d^l_{2\, -2}(\beta) = d^l_{22}(\beta)-
\frac{2(2+\cos\beta)}{\sin^4(\beta/2)} \int_0^\beta \tan^3(\beta'/2)
d^l_{22}(\beta') \, {\rmn{d}}\beta'} \nonumber \\
&& +  \frac{2}{\sin^2(\beta/2)}\int_0^\beta \sec^3(\beta'/2) \sin(\beta'/2)
d^l_{22}(\beta') \, \rmn{d}\beta'.
\label{eq:73}
\end{eqnarray}
Multiplying with $(l+1/2)d^l_{22}(\beta')$ and summing over $l$ we find
\begin{eqnarray}
\xi(\beta) &=& \xi_+(\beta) + \frac{1}{\sin^2(\beta/2)} \int^1_{\cos\beta}
\ud \cos\beta' \, \xi_+(\beta') \sec^4(\beta'/2) \nonumber \\
&-& \frac{2(2+\cos\beta)}{\sin^4(\beta/2)}
\int^1_{\cos\beta} \ud \cos\beta' \, \xi_+(\beta')
\frac{\tan^3(\beta'/2)}{\sin\beta'}.
\label{eq:74}
\end{eqnarray}
As $\xi(\beta)$ depends only on $\xi_+(\beta)$ in the range $(0,\beta)$, it is
possible to construct $\xi(\beta)$ in this range from an unbiased
estimator of $\xi_+(\beta)$ in the same range. By construction,
$\xi(\beta)-\Re \xi_-(\beta)$ will contain only $B$ polarization in the mean.

\begin{figure*}
\epsfig{figure=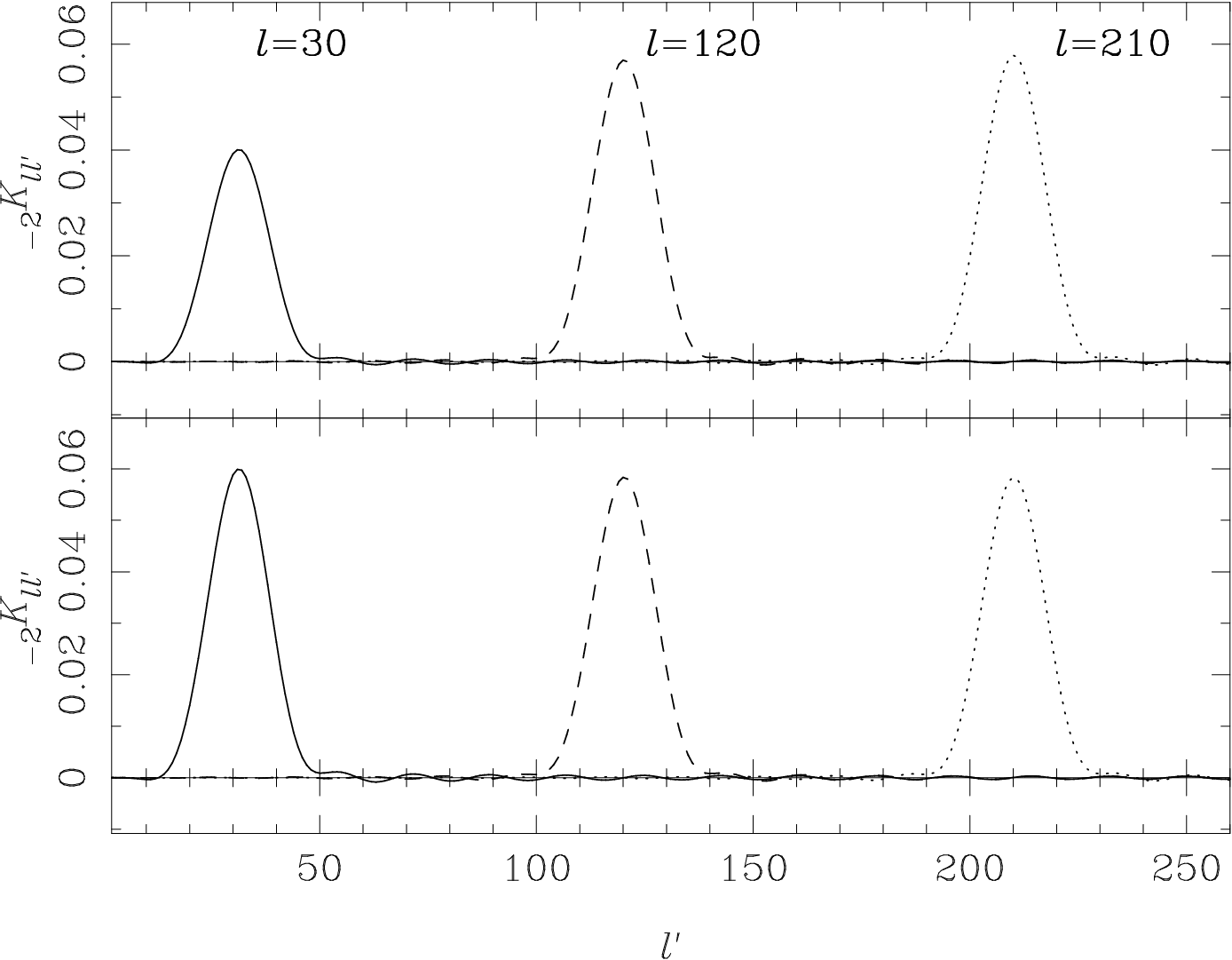,angle=-90,width=8cm}
\epsfig{figure=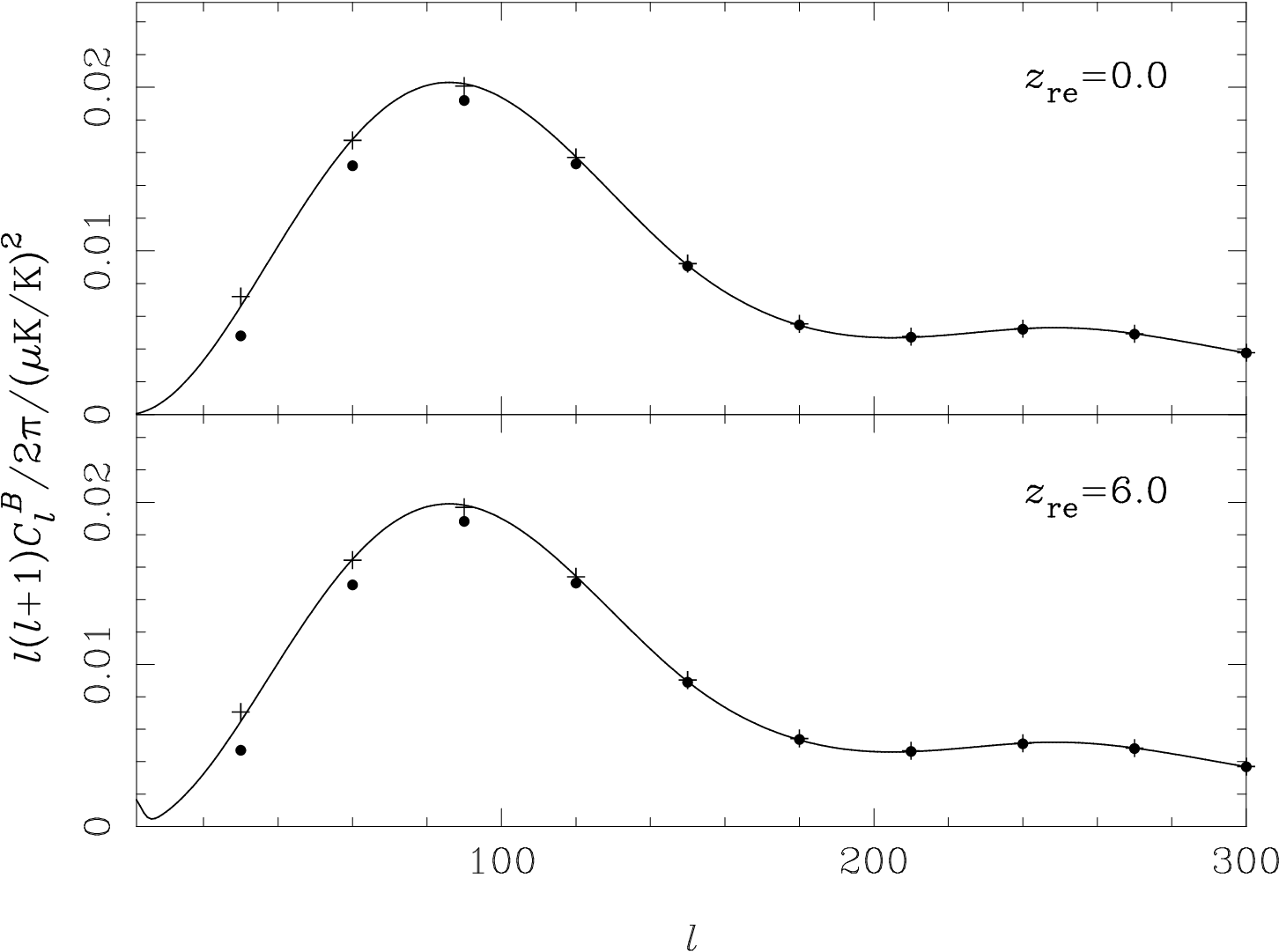,angle=-90,width=8cm}
\caption{Left: window functions ${}_{-2}K_{ll'}$ (top) and their
renormalised counterparts (bottom) for $\beta_\rmn{max}=20^\circ$ and
Gaussian apodization with half-width at half-maximum equal to
$\beta_\rmn{max}/2$. Right: mean recovered $C_l^B$ with (crosses) and
without (circles) renormalisation for the cosmological models
with (bottom) and without (top) reionization.
\label{fig4}}
\end{figure*}

The window function for this method is simply ${}_{-2}K_{ll'}$
(equation~\ref{eq:41}), so that
\begin{equation}
\langle \hat{C}_l^B \rangle = \sum_{l'} {}_{-2}K_{ll'} C_{l'}^B,
\quad
\langle \hat{C}_l^E \rangle = \sum_{l'} {}_{-2}K_{ll'} C_{l'}^E.
\label{eq:75}
\end{equation}
As in the previous section, we can write ${}_{-2}K_{ll'}
= \sum_L {}_{-2}F_{lL}{}_{-2}W_{Ll'}$.
Representative elements of the window functions ${}_{-2}K_{ll'}$
are plotted in Fig.~\ref{fig4} for $\beta_\rmn{max}=20^\circ$ and Gaussian
apodizing with half-width at half-maximum equal to $\beta_\rmn{max}/2$. They
are well approximated by Gaussians with asymptotic normalisation given
by the right-hand side of equation~(\ref{eq:59}). For presentation purposes
it is desirable to have window functions normalised to unity. We can enforce
this by dividing the power spectrum reconstructed from equation~(\ref{eq:69})
by $\sum_{l'}{}_{-2}K_{ll'}$. The exact normalisation is easily computed from
equation~(\ref{eq:48}) by e.g.\ Gauss-Legendre integration, and can be
performed while inverting the correlation functions at negligible computational
cost. The renormalised window functions are also shown in Fig.~\ref{fig4},
along with the mean recovered $C_l^B$, obtained from equation~(\ref{eq:69})
with and without renormalisation by $\sum_{l'}{}_{-2}K_{ll'}$.
The renormalised estimates agree very well in the mean with the true power
spectra. Note also that since we have removed cross-contamination, the
recovered $C_l^B$ are insensitive to large-scale power (from reionization)
in $E$ polarization.

\subsection{Application to BICEP}
\label{bicep}

\begin{figure}
\epsfig{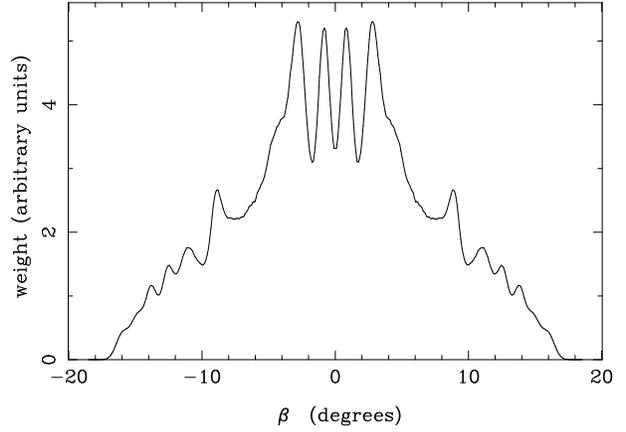}
\caption{Azimuthally-symmetric weight function $w_P(\vnhat)$ adopted for the
BICEP simulations.
We chose to weight in proportion to the integration time per pixel. For white
noise this is equivalent to weighting with the inverse of the noise variance.
\label{figpixwin}}
\end{figure}

\begin{figure*}
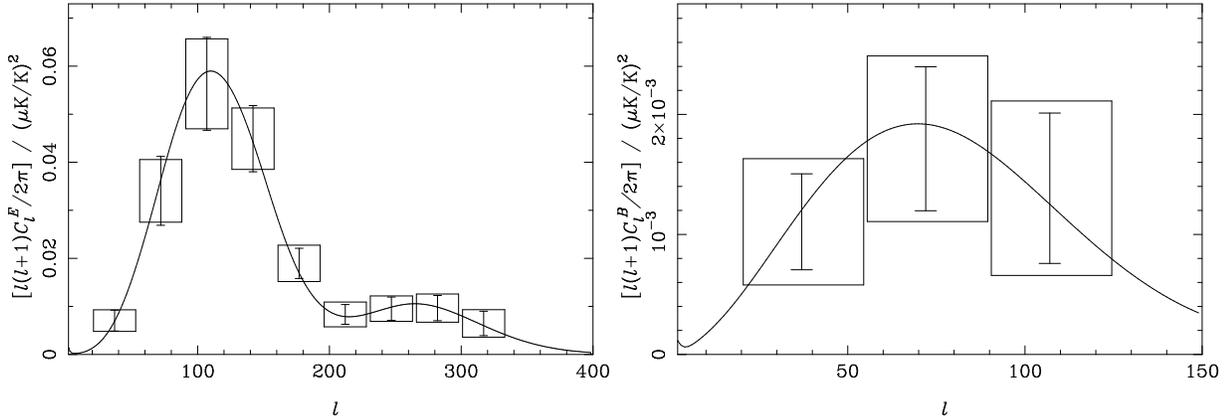

\epsfig{figure=rec_e_bicep.ps,angle=-90,width=8cm}
\epsfig{figure=rec_b_bicep.ps,angle=-90,width=8cm}
\caption{The mean in 100 \textsc{BICEP} simulations of the flat band-powers for
$C_l^E$ (left; smoothed with a 1-degree beam) and $C_l^B$ (right). Monte-Carlo
error estimates from the 100 simulations are shown as boxes, centred on the
average bandpowers from the simulations. The error bars are the
theoretical approximation in equation~(\ref{eq:moderror}). The
solid lines are the input theoretical spectra, smoothed with the beam, for
the cosmological model described in Sec.~\ref{sec:intro} with reionization
at $z=6$. Note they are not convolved with the band-power window function.
\label{figbicep}}
\end{figure*}

As an application of our new estimator, we consider simulated maps for the
\textsc{BICEP} experiment. \textsc{BICEP} is the first of a new generation
of large bolometer arrays that are designed to target $B$-mode polarization.
We used the experimental parameters taken from the \textsc{BICEP}
homepage\footnote{http://bicep.caltech.edu}.
The survey will cover a polar-cap region of angular radius $18\fdg 5$,
integrating for a nominal 300 days. BICEP will be composed of 48
polarization-sensitive bolometers (PSBs) operating at 100 GHz with a
resolution of
$1^\circ$ (full-width at half-maximum) and 48 at 150 GHz with
$0\fdg 7$ resolution. For our simulations we ignored the difference in
beam size between the two channels taking the beam size to be
$1^\circ$, so maps from each channel could be
easily combined without introducing noise correlations. We took each
PSB to have an instantaneous sensitivity of 300 $\mu$K$\sqrt{\rm sec}$.

We simulated 100 noisy CMB maps using a realistic map of the integration time
per pixel based on the BICEP scanning strategy (see Fig.~\ref{figpixwin}).
The cosmology was that described in Sec.~\ref{sec:intro} (with reionization at
$z=6$). The pseudo-$C_l$s were extracted with {\sevensize HEALP}ix at
resolution $N_\rmn{side}=512$ using the weight function shown in
Fig.~\ref{figpixwin}. This corresponds to inverse noise variance weighting
for white noise in the time domain, and the azimuthal symmetry reflects the
symmetry of the proposed BICEP scan strategy.
We generated a further suite of Monte-Carlo noise
realisations which were used to remove the noise bias. The integral in
equation~(\ref{eq:74}) was performed with a cumulative Simpson rule, giving
estimates of $\xi(\beta)$ at the roots 
of a Legendre polynomial scaled to the angular range $(0,31^\circ)$.
In principle we can estimate the correlation functions in the range
$(0,37^\circ)$, but very few pixel pairs contribute to the largest
separation angles so the correlation functions are very noisy there.
During the integration to form $\xi(\beta)$ we constructed $\xi_+(\beta)$
directly from the pseudo-$C_l$s, with the noise bias removed, at points
linearly spaced between the Legendre roots. (We used a nine-point Simpson
rule.)

We recovered the angular power spectra by evaluating equation~(\ref{eq:69})
with Gauss-Legendre quadrature. We adopted a Gaussian apodizing function
with half-width at half-maximum equal to $18\fdg 5$.
We further compressed our estimates into flat band-powers with a
$\Delta l=35$ thus removing much of the sensitivity to the choice of
apodization. We verified with
Monte-Carlo simulations that $\Delta l=35$ is sufficient to remove
any significant correlations between adjacent band powers.
The mean band-powers for $C_l^E$
and $C_l^B$ (smoothed with the $1^\circ$ beam) from 100 simulations are
plotted in Fig.~\ref{figbicep}, along
with $\pm \sigma$ error boxes estimated from the simulations. From the
simulations we have verified that the method is unbiased (to within
the standard error of the Monte-Carlo averages).
%
We also compared the errors estimated from the 100 simulations
with the rule of thumb in equation~(\ref{eq:error}). We found that to get
good agreement with the Monte-Carlo errors on $C_l^E$ it was necessary to
refine equation~(\ref{eq:error}) to take account of the noise
inhomogeneity and the compression to band powers more properly.
We used the following approximation to the covariance of the
recovered beam-smoothed $C_l^E$s (see~\citealt{efstathiou03} for a
derivation of this formula for the temperature anisotropies):
\begin{eqnarray}
\rmn{cov}(\hat{C}_l^E, \hat{C}_{l'}^E) &\approx&
\frac{1}{2\pi(w_2 f_{\rmn{sky}})^2}
\sum_L \Big[C_l^E C_{l'}^E \sum_M |(w_P^2)_{LM}|^2 \nonumber \\
&+& 2 \sqrt{C_l^E C_{l'}^E}
\sum_M (w_P^2 \sigma_P^2)_{LM} (w_P^2)_{LM}^\ast \nonumber \\
&+&  \sum_M | (w_P^2 \sigma_P^2)_{LM} |^2 \Big]
\left( \begin{array}{ccc} l & l' & L \\ 0 & 0 & 0 \end{array} \right)^2.
\label{eq:moderror}
\end{eqnarray}
Here, $\sigma_P^2(\vnhat)$ is the polarization noise variance per solid
angle (see equation~\ref{eq:34}), $w_P(\vnhat)$ is the polarization weight
function, and e.g.\ $(w_P^2 \sigma_P^2)_{LM}$ are the (spin-0) multipoles
of the product $w_P^2(\vnhat) \sigma_P^2(\vnhat)$. Equation~(\ref{eq:moderror})
makes a number of approximations: (i) it takes no account of the need to
separate $E$ and $B$, which is reasonable for $E$ given that it dominates
$B$, but will not be valid if extended to $B$ when
$C_l^E / C_l^B > (l \Delta_w)^2$,
where $\Delta_w$ is the characteristic width of $w_{P,l}$;
(ii) it ignores the spin-2 nature of the polarization, which is acceptable
at high $l$; and (iii) it only treats the inversion from
pseudo-$C_l$s to $\hat{C}_l$s approximately (i.e.\ divide by $w_2
f_{\rmn{sky}}$). We defer a full discussion of analytic approximations to
the covariance of polarization power spectra to a future paper (Challinor
\& Chon in preparation), where we show how to generalise
equation~(\ref{eq:moderror}) to take account of $E$-$B$ mixing on an apodized
sky. Here, we simply note that the above assumptions should hold well
for $C_l^E$ in this application to BICEP. For the level of $C_l^B$ in
our assumed cosmology ($r=0.31$), and given the smooth apodization of the
edges of the survey region by $w_P(\vnhat)$, the application of
equation~(\ref{eq:moderror}) to $B$ should serve as a useful first
approximation to the errors. Note also that, for uniform noise, we recover 
equation~(\ref{eq:error}) if we average into bands that are wide
compared to the power spectrum of the square of the weight function, 
$\sum_m |(w_P^2)_{lm}|^2 / (2l+1)$. The theoretical approximation
to the errors in Fig.~\ref{figbicep} are obtained by
summing equation~(\ref{eq:moderror}) over $l$ and $l'$ in a given band.
These theoretical predictions agree well with the Monte-Carlo errors
for $E$, and are in broad agreement for $B$. As expected, in the latter case
the details of the $E$-$B$ separation process that are ignored in our
rough theoretical predictions are more critical.

\begin{figure}
\epsfig{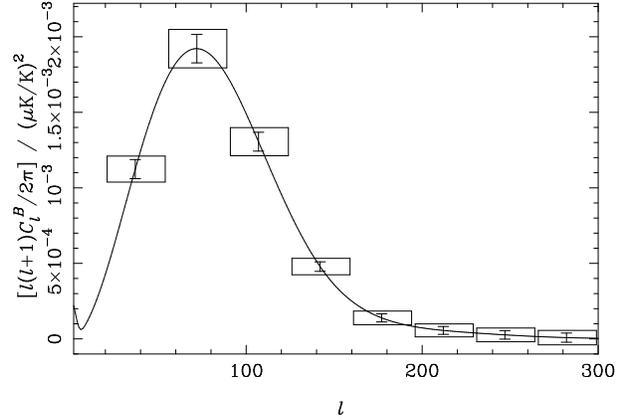}
\caption{The variance of an estimate of $C_l^B$ obtained by 
equation~(\ref{eq:newmoderror}) in the absence of noise with the BICEP
sky coverage. The boxes are the full error, including the cross-contribution
from $E$ modes. The bars are the variance obtained by setting
$C_l^E$ to zero, and thus are representative of the errors that would
be achieved if $B$ were separated from $E$ at the level of the map.
\label{figerror2}}
\end{figure}

Our new estimator removes the cross-contamination between $E$
and $B$ in the mean, however, the variance of an estimate
of $C_l^{E}$ or $C_l^B$ contains a contribution from both $E$ and $B$ modes.
To assess more carefully the level of cross-contamination due to the
geometric effect of $E$-$B$ mixing, we compute exactly the covariance
of our decoupled power spectrum estimates in the absence of noise. Since the
cross-contamination will be more significant for $B$ than $E$, we concentrate
on the former. We compute the error covariances first with $E$ and $B$ power
retained, and then with only the $B$ power. The latter calculation
approximates the errors
we would obtain if we separated the $E$ and $B$ modes at the level of the map
prior to power spectrum estimation. The mechanics of the calculation are
as follows. First, we compute the covariance of the polarization
pseudo-$C_l$s using the techniques described by~\citet{hansen02b}. This is
only tractable beacuse of the azimuthal symmetry of the BICEP scanning
strategy. We then linearly transform the pseudo-$C_l$ covariance to that
of the decoupled estimates using the fact that our power spectrum estimation
method is linear in the pseudo-$C_l$s, i.e.\
\begin{equation}
{\rm{cov}}(\hat{C}_l^X,\hat{C}_{l'}^Y)
= \sum_{L,L',X',Y'} M_{lL}^{XX'} {\rm{cov}} (\tilde{C}_L^{X'}, 
\tilde{C}_{L'}^{Y'}) M_{l'L'}^{YY'},
\label{eq:newmoderror}
\end{equation}
where $X$, $X'$, $Y$ and $Y'$ run over $E$ and $B$, and the
coupling matrices relate the decoupled power spectrum estimates to the
pseudo-$C_l$s:
\begin{equation}
\hat{C}_l^X = \sum_{l' X'}M_{ll'}^{XX'} \tilde{C}_{l'}^{X'}.
\end{equation}
The columns of the coupling matrices $M_{ll'}^{XX'}$ are conveniently
extracted by setting all of the pseudo-$C_l$s to zero except for
$\tilde{C}_{l'}^{X'}$ in our power spectrum code, and then reading out
the recovered $\hat{C}_l^X$. The matrix is symmetric on the indices $X$ and
$X'$. Figure~\ref{figerror2} summarises our results obtained for the
$B$-mode power spectrum. As in Fig.~\ref{figbicep} we compress our
results into flat band-powers. We see that, in the noise-free case, the
errors we obtain using the BICEP weight function $w_P(\vnhat)$
are not dominated by the cross-contribution from $E$-mode polarization;
this accounts for approximately only 20-per cent of the error budget.
Of course,
a lower level of $B$-mode polarization would increase the relative
significance of the cross-contribution. Ultimately, this cross-contamination
may make our method unsuitable for future, high-sensitivity $B$-mode
experiments surveying small regions if the tensor amplitude is too low.
We leave quantification of this statement to a future paper (Challinor
\& Chon in preparation), where we show how to estimate the cross-contribution
to the variance for non-symmetric weight functions.

Since full covariance information is available from
equation~(\ref{eq:newmoderror}), the level of correlation between adjacent
band-powers can be calculated directly. For the BICEP scan
strategy, $M_{ll'}^{XY}$ oscillates with a full-width at half-maximum
of approximately 12. Hence, for the chosen bin width $\Delta l=35$, the
correlation between adjacent bins is negligible.

\section{Conclusion}
\label{conclusion}

We presented a fast and unbiased method to extract CMB polarization power
spectra from large maps via the two-point correlation functions.
The method, which generalises that of~\citet{szapudi01} to polarization,
can be summarised as follows. First, we compute unbiased
estimates of the three (complex) polarization auto- and cross-correlation
functions at the roots of a Legendre polynomial from pseudo-$C_l$s of
heuristically-weighted maps. The estimates
of the correlation functions can be computed in $O(N_\rmn{pix}^{3/2})$
operations using fast spherical transforms. If the correlation functions can be
estimated for all angular separations, the power spectra can be accurately
recovered with Gauss-Legendre integration. In this case, the method is
unbiased: the theoretical window function is a Kronecker delta. Further
compression to band-powers can then be made, and the resulting theoretical
window functions would be top-hat functions. If the correlation functions
cannot be estimated for all angular separations, due to limitations of
sky coverage, we showed that significant $E$-$B$ mixing can occur.
In particular, large-scale $E$ power (due to
reionization) can be aliased into $B$ on all scales. Although $E$-$B$
mixing does not present a fundamental problem for parameter
extraction, it does complicate the interpretation (and presentation) of the
recovered power spectra. For this reason, we proposed a new estimator,
extending earlier work by~\citet{crittenden02}, that removes $E$-$B$ mixing
exactly in the mean when working with incomplete correlation functions.
Note that this is not the case for regularised inversions of the pseudo-$C_l$s
in harmonic space (e.g.\ by working with pseudo band powers as in a polarized
extension of MASTER; \citealt{hivon02}).
The new estimator requires one further numerical integration of the estimated
correlation function $\hat{\xi}_+(\psi)$ to obtain functions that
contain only $E$ or $B$ power in the mean. Using e.g.\ a cumulative Simpson
rule, these functions can be estimated accurately at the roots of a Legendre
polynomial and inverted to power spectra with Gauss-Legendre quadrature.
The increase in computational effort is minimal, and the
theoretical window functions that result do not couple $E$ and $B$ power
in the mean by construction. Fourier ringing in the estimates can be
safely controlled by apodizing the integral transforms (or by compressing
into band-powers), without introducing any $E$-$B$ mixing. 

An essential part of our method (and indeed any quadratic method) is
being able to remove the mean noise contribution from the correlation
functions
(i.e.\ to remove the bias due to the noise). For general noise properties
we must resort to Monte-Carlo evaluation of the mean over an ensemble of pure
noise realisations. The method presented here is thus dependent on being able
to simulate noise maps efficiently. 
Error estimation on the recovered $C_l$s must also generally proceed by
Monte-Carlo evaluation. The $O(N_\rmn{pix}^{3/2})$ scaling of our method
makes this a realistic proposition, even for mega-pixel maps.

We applied our methods to simulations of a large-area survey, with
parameters similar to \emph{Planck}, and also to the BICEP experiment
which will cover 3 per cent of the sky. In both cases we obtained errors
in line with theoretical expectations.
Although our algorithm in its present
form is already practical for analysing mega-pixel CMB maps, further work is
required to assess the optimality of the underlying methods.
In particular, comparison with current (brute-force)
maximum-likelihood codes should be possible for low-resolution simulations;
comparison at higher resolution must await further algorithmic
development of the likelihood codes. Another issue worth investigating
further is the impact of the choice of pixel weighting
on the cosmic variance contribution from e.g.\ $C_l^E$ to the recovered
$C_l^B$. Although we have separated $E$ and $B$ in the mean, this
does not guarantee that in a single realisation there is no recovered
$B$ power due to $E$ modes in the signal. We have shown that this geometric
effect does not dominate the error budget for $C_l^B$ for the BICEP weight
function, in the absence of noise, for a tensor-to-scalar ratio of $r=0.31$.
However, ultimately this cross-contribution may limit the applicability of
our method if the tensor amplitude is low enough (Challinor \& Chon in
preparation). To eliminate this undesirable contribution
to the variance would require separating $E$ and $B$
at the level of the map, e.g.~\citet{lewis02}, prior to performing power
spectrum estimation.

\section*{Acknowledgements}

We thank Robert Crittenden for useful discussions at the start of this
project, Antony Lewis for help with the integrals in the Appendix,
and Stephane Colombi for help with issues of implementation.
ADC acknowledges a Royal Society University Research Fellowship, and IS was
supported by NASA through AISR grants NAG5-10750, NAG5-11996, and ATP grant
NASA NAG5-12101 as well as by NSF AST02-06243. Some of the results in this
paper have been derived using the {\sevensize HEALP}ix~\citep{gorski} package.

\appendix
\section{$A(\psi)$ for uniformally-weighted,
azimuthally-symmetric patches}

In the special case that the analysis is performed over an
azimuthally-symmetric part of the sky with uniform pixel weighting
$w(\vnhat)=1$, the correlation function normalisation $A(\psi)$ can be
evaluated analytically. (We can drop the subscripts $P$, $T$ and $X$ in this
case since there is no distinction between the normalisations.)

We begin by considering a polar-cap region with angular radius $\alpha$,
and assume that $\alpha \le \pi/2$. The integral in equation~(\ref{eq:21})
then evaluates to give
\begin{eqnarray}
\frac{1}{A(\psi)} &=& 4\pi \left[
\cos^{-1}\left(\frac{2\sin^2(\psi/2)}{\sin^2\alpha}-1 \right) \right.
\nonumber \\
&& \mbox{} \left. -\cos\alpha\cos^{-1}\left(\frac{2\tan^2(\psi/2)}{\tan^2
\alpha}-1\right) \right]
\label{eq:app1}
\end{eqnarray}
for $\psi \le 2\alpha$, and zero otherwise. Note that $1/A(0) = 4\pi^2
(1-\cos\alpha) = 8\pi^2 f_{\rmn{sky}}$ as required by
equation~(\ref{eq:38}). Note also that $1/A(\psi)$ goes continuously to
zero as $\psi \rightarrow 2\alpha$. We can now construct the other important
case of a Galactic cut by symmetry. For a symmetric cut
subtending an angle $\alpha_\rmn{c}$, provided that $\alpha_\rmn{c}
\le \pi/2$, there are pixel pairs at all angular separations and
$1/A(\psi)$ is non-zero for all $\psi \in (0,\pi)$. If we denote
the normalisation for a polar-cap region of angular radius
$(\pi-\alpha_\rmn{c})/2$ by $A_\ast(\psi)$, then we find
\begin{equation}
\frac{1}{A(\psi)} = \left\{ \begin{array}{ll}
\frac{2}{A_\ast(\psi)} & 0\le \psi \le \alpha_\rmn{c}, \\
\frac{2}{A_\ast(\psi)} + \frac{2}{A_\ast(\pi-\psi)} &
\alpha_\rmn{c} \le \psi \le \pi - \alpha_\rmn{c}, \\
\frac{2}{A_\ast(\pi-\psi)} & \pi -\alpha_\rmn{c} \le
\psi \le \pi.
\end{array} \right.
\label{eq:app2}
\end{equation}

\bsp  
\label{lastpage}
\end{document}